\documentclass[10pt]{article}
\usepackage[letterpaper, margin=1in]{geometry}
\usepackage{fancyhdr}
\usepackage{tabularx}
\usepackage{multirow}
\usepackage{booktabs}
\usepackage{longtable}
\usepackage{hyperref}
\usepackage{adjustbox}
\usepackage{verbatim}
\usepackage{booktabs}
\usepackage{subcaption}
\usepackage{authblk}
\usepackage{titlesec}

\usepackage{amsmath}
\usepackage{amsfonts}
\usepackage{graphicx} 
\usepackage[backend=biber, style=numeric,sorting=none]{biblatex}
\title{WGTDA: A Topological Perspective to Biomarker Discovery in Gene Expression Data}

\author{Ndivhuwo Nyase $^{1}$, Lebohang Mashatola $^{1}$, Aviwe Kohlakala $^{1}$, Kahn Rhrissorrakrai $^{2}$, Stephanie Müller}

\small \affil[1]{IBM Research Africa, Johannesburg, South Africa}
\small \affil[2]{IBM Research, Yorktown Heights, NY, USA}
\date{}

\begin{document}


\maketitle

\begin{abstract}
\noindent



Advancing the discovery of prognostic cancer biomarkers is crucial for comprehending disease mechanisms, refining treatment plans, and improving patient outcomes. This study introduces Weighted Gene Topological Data Analysis (WGTDA), an innovative framework utilizing topological principles to identify gene interactions and distinctive biomarker features. WGTDA undergoes evaluation against Weighted Gene Co-expression Network Analysis (WGCNA), underscoring that topology-based biomarkers offer more reliable predictors of survival probability than WGCNA's hub genes. Furthermore, WGTDA identifies gene signatures that are significant to survival probability, irrespective of whether the expression is above or below the median. WGTDA provides a new perspective on biomarker discovery, uncovering intricate gene-to-gene relationships often overlooked by conventional correlation-based analyses, emphasizing the potential advantage of leveraging topological concepts to extract crucial information about gene-gene interactions. \\



\end{abstract}
\newpage
\section{Introduction}

Omics studies offer the ability to unravel complex interactions within cellular and molecular systems that drive disease processes, and are thus pivotal in biological research for identifying prognostic biomarkers for complex diseases such as cancer \cite{jeong2023current}. Biomarkers are critical in spearheading drug development and bridging molecular understanding with clinical practice offering a pathway to earlier and more accurate disease diagnoses, improved understanding of disease mechanisms, and overall improvement in patient care \cite{oldenhuis2008prognostic}. Significant challenges hindering biomarker discovery includes the high-dimensionality of omics data, the ability to discern signal from noise, the validation of biomarkers across diverse populations and conditions, and the implementation of analytical methods to detect subtle but clinically relevant patterns within the data \cite{mcdermott2013challenges}. These challenges demonstrate the need for developing novel methods for biomarker discovery, presenting a unique opportunity for exploring Topological Data Analysis (TDA) as a means of genomic data exploration, thus potentially improving our understanding of disease mechanisms and improving patient outcomes. TDA is a set of computational topology techniques used to uncover the intricate local and global topological structures hidden within data, thus offering a unique perspective for omics data analysis \cite{chazal2021introduction}. \\

Here, we present a novel framework, Weighted Gene TDA (WGTDA), implemented to identify novel biomarkers in gene expression data. WGTDA provides a topological perspective on the organization of gene expression networks, further deciphering the complexity embedded in sequence-based gene expression data. In this study, WGTDA is compared to Weighted Gene Co-expression Network Analysis (WGCNA), a data mining technique widely used for identifying modules of co-expressed genes and hub genes within these modules \cite{langfelder2008wgcna}. Contrary to the principles of correlation and hierarchical clustering used by WGCNA, WGTDA is adept at uncovering intricate patterns, holes, and structures that may be overlooked by traditional correlation-based methods\cite{edelsbrunner2008persistent}. Furthermore, functional enrichment and survival analyses on the identified gene signatures of both frameworks were conducted to validate and establish the clinical relevance and utility of WGTDA. To provide a comprehensive comparison, Breast Cancer (BRCA), Lung Adenocarcinoma (LUAD), and Colorectal Adenocarcinoma (COAD/READ) data from The Cancer Genome Atlas (TCGA) were utilized in this exploration.  \\

Survival analyses revealed that WGTDA was able to identify unique gene signatures that are correlated and important to survival probability regardless of whether the expression were above or below the median. These gene interactions indicate that the complex underlying behaviour may not be influenced by gene expression alone. Moreover, WGTDA proved to be more effective than WGCNA in uncovering prognostic biomarkers that better predict survival outcomes. Notably, WGTDA pinpointed gene signatures that are known to contribute specifically to lung and breast cancer, underscoring its potential in targeted cancer research. Such signatures can further elucidate the mechanisms governing complex diseases to develop effective interventions and enhance cancer patient outcomes.

\section{Background}

\subsection{Correlations-Based Networks}

Correlation-based network analyses, including WGCNA, have notable limitations that influence their interpretability. One significant challenge is the inability to discern causal relationships from correlations, leading to potential misinterpretations of network structures \cite{ness2016correlation}. Correlation does not imply causation, and spurious correlations may arise, giving rise to misleading conclusions about the strength or nature of connections between gene-interactions \cite{calude2017deluge}. Additionally, correlation-based approaches may overlook non-linear relationships, limiting their ability to capture the multi-dimensional nature of gene interactions accurately, potentially overlooking critical insights offered by more intricate patterns in the data \cite{wang2011study}. Furthermore, WGCNA uses hierarchical clustering and a dynamic tree cutting approach to define gene modules. Determining the optimal number of clusters or modules remains a challenge, highlighting the need for novel biomarker discovery methods\cite{langfelder2008wgcna}. TDA offers a novel perspective by analyzing the shape of data, revealing hidden structures and patterns not found using other methods. To our current knowledge there are no biomarker discovery tools built on the principles of algebraic topology.

\subsection{Topological Data Analysis}

Topology is the mathematical study of shapes and spatial properties in data that remain invariant to continuous deformations.
TDA leverages the principles of algebraic topology to deduce and examine the intricate structures inherent in complex datasets. It involves the representation of data points as a simplicial complex, where the relationships between these points are determined by correlations or distances within a topological space \cite{lum2013extracting}. The simplicial complex can be defined by $X$, described as a finite metric space, and $K$, the set of simplicial complexes. Here, $K_i$ is the simplicial complex at filtration step $i$:

 \[
 X \hookrightarrow K_0 \hookrightarrow K_1 \hookrightarrow \ldots \hookrightarrow K_n
 \]

A simplicial complex is a construct encompassing an array of interconnected elements, ranging from points to line segments, triangles, and their n-counterparts. Here, we utilize a Vietoris-Rips (VR) complex which is a simplicial complex where its k-simplices are defined by subsets of $(k + 1)$ points within a set $X$, each having a diameter that does not exceed a specified $\epsilon$ threshold. \\

\begin{figure}[htp!]
     \centering
     \includegraphics[width=\textwidth]{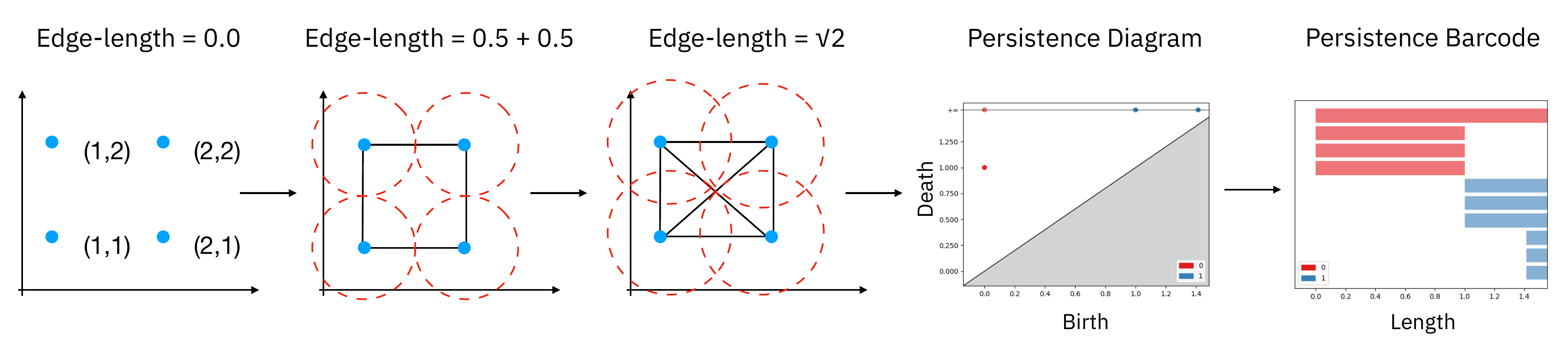}
     \caption{Iterative growth of the simplicial complex achieved by increasing the radius around each point subsequently forming the connection of edges. Followed by a persistence diagram and persistent barcode summarising the length of the topological feature colored by specific \textit{Betti} number. }
     \label{fig:summary_workflow}
 \end{figure}

The identification of topological features using TDA is centered around persistent homology, a mathematical tool that employs persistence-based filtration. In persistent homology, a filtration process is applied to simplicial complexes using growing balls around each data point. As these balls expand, the intersection of growing balls between adjacent data points results in the formation of edges, connecting the data points (Figure \ref{fig:summary_workflow}). This process continues, creating higher-dimensional simplices such as triangles and tetrahedra as the filtration progresses~\cite{wasserman2018topological}. \\ 

The non-linear relationships captured by persistent homology can manifest as connected points, loops, voids, or other complex structures in the simplicial complex and are categorized by \textit{Betti} numbers.

\subsection{Betti Number}
\textit{Betti} numbers are used to differentiate topological spaces based on the connectivity of \textit{n}-dimensional simplicial complexes. For example, \textit{Betti}-0 corresponds to the number of connected components or clusters, \textit{Betti}-1 represents the number of non-contractible loops or cycles, and \textit{Betti}-2 indicates the number of voids or enclosed regions in the data space \cite{xia2015multidimensional}. These structures provide a more comprehensive understanding of the data's intrinsic topology, revealing hidden patterns and relationships. Furthermore, the \textit{Betti} numbers are encoded on the persistence diagram and persistence barcode, which tracks the persistence of features by recording birth (b) and death scales (d) and can be defined as: $(b, d) \mid b, d \in \mathbb{R}, 0 \leq b \leq d < \infty$. Each point $(b, d)$ represents the birth and death scales of a topological feature \cite{pranav2017topology}.

\subsection{Related Work}
Recent studies have shown that the application of machine learning techniques to a set of topological representations in transcriptomic data yielded promising results in various classification tasks \cite{mandal2020topological,dey2021gene}. These studies underscore the significance of identifying topological features as dependable indicators for detecting the presence of a disease, effectively unveiling informative signals embedded in high-dimensional genomic datasets. In a related study, notable distinctions have been observed in topological summaries when comparing normal- and cancerous samples \cite{masoomy2021topological}. The examination of \textit{Betti} curves for cancer samples suggests a notable biological phenomenon related to oncogene addiction at a network level. These findings contribute to the growing body of evidence, emphasizing the role of topological features in discerning critical distinctions between normal and pathological conditions, particularly within complex trascriptomic datasets \cite{masoomy2021topological}. This study contributes to the scientific body of knowledge by introducing a novel technique utilizing topology for biomarker discovery.

\section{Methods} 

\subsection{Data selection and preprocessing}

For a comprehensive exploration of biological variations associated with adenocarcinomas in different organs, gene expression datasets from TCGA (available at: \url{https://portal.gdc.cancer.gov}) were obtained \cite{weinstein2013cancer}. Three datasets for common cancers were selected, namely Breast Cancer (BRCA), Lung Adenocarcinoma (LUAD), and Colorectal Adenocarcinoma (COAD/READ). Each dataset comprised of RNA-Sequencing (RNA-Seq) data, thereby ensuring a high-resolution view of gene expression and facilitating a reliable exploration of the intricate molecular features associated with the mechanisms governing the BRCA, COAD/READ, and LUAD cohorts.
\\



A set of 326 cancer-related genes were extracted from the gene expression data using information from the Kyoto Encyclopedia of Genes and Genomes (KEGG) database, available at\url{https://www.genome.jp/kegg/pathway.html} \cite{kanehisa2007kegg}. These genes play crucial roles in regulating multiple signaling pathways, including Extracellular Signal-Regulated Kinase (ERK), Phosphoinositide 3-Kinase (PI3K), Rat Sarcoma (RAS), Wingless/Integrated (WNT), Neurogenic Locus Notch Homolog (NOTCH), Hedgehog (HH), Calcium, Hypoxia-Inducible Factor 1 (HIF-1), Nuclear reception, Kelch-like ECH-Associated Protein 1 - Nuclear Factor Erythroid 2-Related Factor 2 (KEAP1-NRF2), cell cycle, apoptosis, and telomerase activity. Subsetting genes narrowed the vast genomic search space for better understanding of the genetic mechanisms of various cancer forms. Moreover, this reduced the computational explosion of obtaining the topological features using high dimensional datasets. Furthermore, FPKM (Fragments Per Kilobase Million) was employed to mitigate biases associated with variations in gene length \cite{love2016modeling}. 





\subsection{WGCNA Framework}
WGCNA was employed to identify co-expressed gene modules in each of the cancer datasets. First, a pairwise distance correlation matrix was computed from the gene expression data, capturing the statistical relationships between each pair of genes and providing a comprehensive view of their co-expression patterns. Thereafter, complete linkage hierarchical clustering was performed to identify gene modules.  The top 15\% most connected genes within gene modules (i.e., hub genes) were determined using \textit{kME} (eigengene-based connectivity) as a thresholding metric as described elsewhere\cite{bi2015gene,shalgi2007global,xu2022differential,fuller2007weighted,wang2015identification}:

\begin{equation}
    kME(i, M) = \frac{\text{cov}(X_i, X_M)}{\sqrt{\text{var}(X_i) \cdot \text{var}(X_M)}}
\end{equation}

Where $\text{cov}(X_i, X_M)$ is the covariance between the expression profile of gene $i$ (denoted as $X_i$) and the module eigengene $X_M$, $\text{var}(X_i)$ is the variance of the expression profile of gene $i$, and $\text{var}(X_M)$ is the variance of the module eigengene. A higher \textit{kME} value indicates a stronger correlation between the gene and the module eigengene, signifying the genes role as a hub gene within the network \cite{zhang2005general}. Hub genes are pivotal in gene co-expression networks, significantly influencing the network structure. They serve as crucial biomarkers, representing collective expression dynamics \cite{nguyen2021identification, guo2020identification, li2020weighted}.

\subsection{WGTDA Framework}

\begin{figure}[h]
    \centering
    \includegraphics[width=\textwidth]{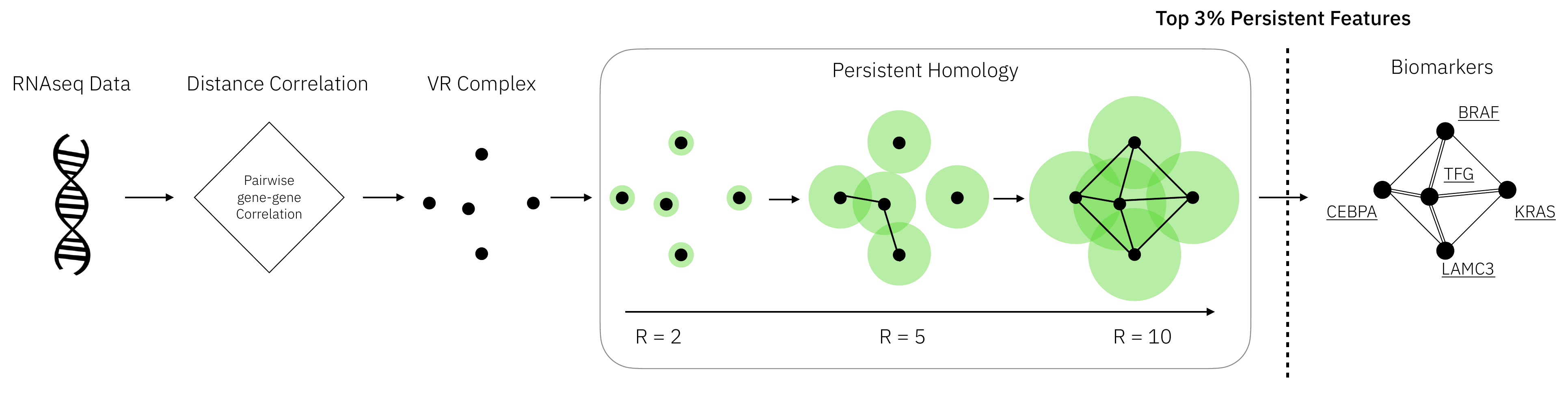}
    \caption{Illustration of the step-by-step process of the WGTDA, starting from data acquisition and preprocessing, through to the identification of topological features and the subsequent analysis for biomarker discovery.}
    \label{fig:WGTDA-Framework}
\end{figure}

Here, we propose a novel framework, WGTDA, to identify potential biomarkers predictive of survival outcomes (Figure \ref{fig:WGTDA-Framework}). A pairwise distance correlation comparable to the distance matrix utilized for WGCNA was computed for the WGTDA analysis. This approach facilitated a consistent and valid comparison for both WGTDA and WGCNA to identify clusters of genes with the same co-expression patterns. Subsequently, a VR complex was constructed from the distance matrix. Each gene represented in the distance matrix is analogous to a 0-simplex and each interaction between genes are conceptualized as \textit{n}-dimensional simplices, forming part of the VR complex. \\

Following the construction of the VR complex, persistent homology was employed to compute the persistent topological features or the \textit{Betti} numbers within the complex. The top 3\% of persisting features for \textit{Betti}-1 and \textit{Betti}-2 were selected with the aim of mitigating noise and highlighting more reliable topological patterns \cite{pranav2017topology, chung2018topological}. Furthermore, we adjusted the \% persistence threshold to ensure a comparable number of genes to compare and assess between the two methodologies. The identified topological features underwent \textit{in-silico} validation, a process crucial for confirming their potential as reliable biomarkers and highlighting prospective utility in predicting clinical outcomes.

\section{Survival Analysis}

\subsection{Kaplan-Meier Survival Analysis}

In this study, a pivotal aspect was the exploration of the prognostic significance of identified biomarkers from WGCNA and WGTDA through survival analysis. The identified hub genes and persistent \textit{Betti} features underwent \textit{in silico} validation via Kaplan-Meier survival analysis. This involved leveraging survival data from the BRCA, COAD/READ and LUAD cohorts and stratifying patients based on the median expression levels of the proposed gene signatures. This median-based stratification served as a threshold to distinguish between 'high' and 'low' expression groups. The 'high' group encompassed patients whose signature expression levels were above the median, indicating a potentially heightened biological activity. In contrast, the 'low' group included those with expression levels below the median, suggesting a reduced activity. The primary objective of this stratification was to analyze and compare the survival outcomes between the 'high' and 'low' expression groups with respect to the proposed biomarker discovery technique. The $p$-values from the log-rank tests, were reported to quantify the magnitude and significance of the observed differences \cite{goel2010understanding}. 




\subsection{Random Survival Forest}
A Random Survival Forest (RSF) analysis was conducted to assess the prognostic potential of hub genes and \textit{Betti}-1/\textit{Betti}-2 gene signatures. In contrast to the Kaplan-Meier approach, the RSF model was not based on prior stratification of expression levels. The RSF approach takes in the specific gene signature expression levels for all patients and assesses each technique's contribution to survival probability. This approach allowed for a more nuanced understanding of how biomarkers influence survival outcomes. A critical aspect of RSF analysis was the evaluation of each biomarker's predictive power in relation to patient survival. This was done with variable importance which is a statistical concept used to rank the relative significance of different variables in their contribution to the predictive power of a model \cite{gregorutti2017correlation}. Here, variable importance was used as a metric for determining how important a particular gene signature is in predicting survival. All trees were run to 1,000 iterations, employing the log-rank splitting rule to optimize tree growth~\cite{beyene2009determining}. Upon identifying important gene signatures, functional enrichment was performed for biological interpretation.

\section{Functional Enrichment}
Significant gene signatures pivotal for survival probability were identified and subsequently subjected to functional enrichment analysis using the Reactome pathway knowledgebase accessed using the R programming language tool ClusterProfiler \cite{wu2021clusterprofiler}. The Reactome pathway knowledgebase utilizes a hypergeometric distribution test to assess for over-representation of biological pathways within a given gene list \cite{jassal2020reactome}. This method identifies significant pathways that are disproportionately represented, offering insights into the biological functions and processes associated with the hub genes and the topological features under study. Statistical validation of the enriched biological pathways was conducted using a 5\% False Discovery Rate (FDR) Benjamini-Hochberg (BH) adjusted $p$-value threshold to limit uncertainty \cite{mubeen2019impact}.

\section{Results}

\subsection{Data Selection and Preprocessing}

\begin{figure}[!htb]
    \centering
    \includegraphics[width=0.42\textwidth]{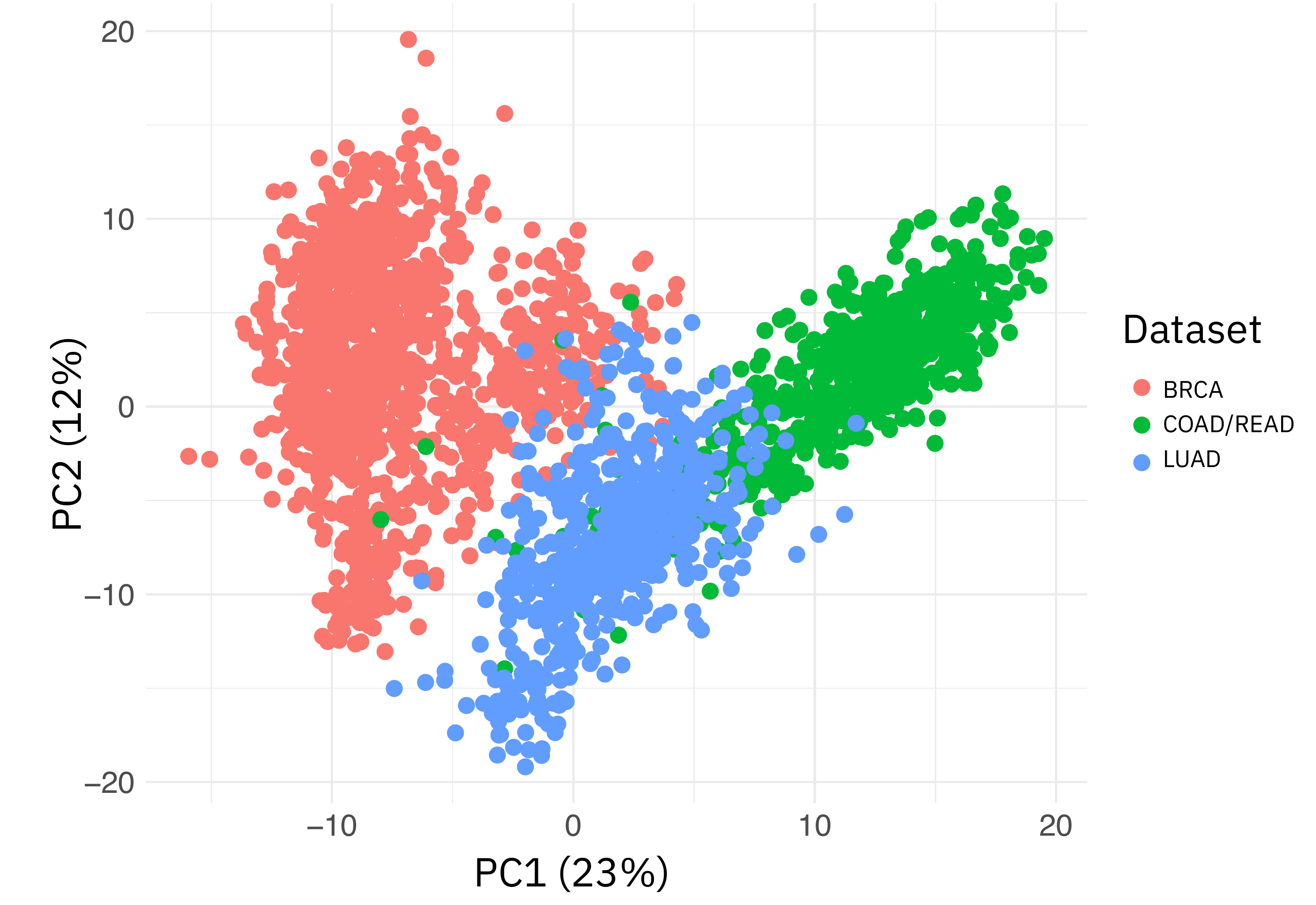}
    \caption{PCA plot representing the distribution of gene expression for TCGA BRCA, COAD/READ and LUAD datasets in a reduced-dimensional space. Each point and its position on the plot corresponds to a cancer patient with regard to the first two principal components (PC1 and PC2).}
    \label{fig:pca-plots}
\end{figure}

For the WGTDA and WGCNA analyses, the datasets comprised of a total of 193, 124, and 132 patients for BRCA, COAD/READ and LUAD, respectively. Moreover, the datasets used for the survival analyses consisted of a total of 1230, 695 and 600 patients for BRCA, COAD/READ and LUAD. The purpose of a smaller dataset for the WGCNA and WGTDA analyses was to ensure robust validation on a larger, independent dataset. This approach enabled the training and testing datasets to remain independent, enhancing the confidence of the study's findings.  
Principal Component Analysis (PCA) was performed to visualize the degree of separation of transcriptomic profiles for three distinct cancer types (Figure \ref{fig:pca-plots}). A clear distinction and variability between cancer types is observed when utilizing the same pre-selected gene sets to define gene expression matrices.


\subsection{Gene Signature Identification using WGCNA and WGTDA}

\subsubsection{WGCNA}

\begin{figure}[htp!]
\centering

  \begin{subfigure}{0.32\textwidth}
    \includegraphics[width=\linewidth]{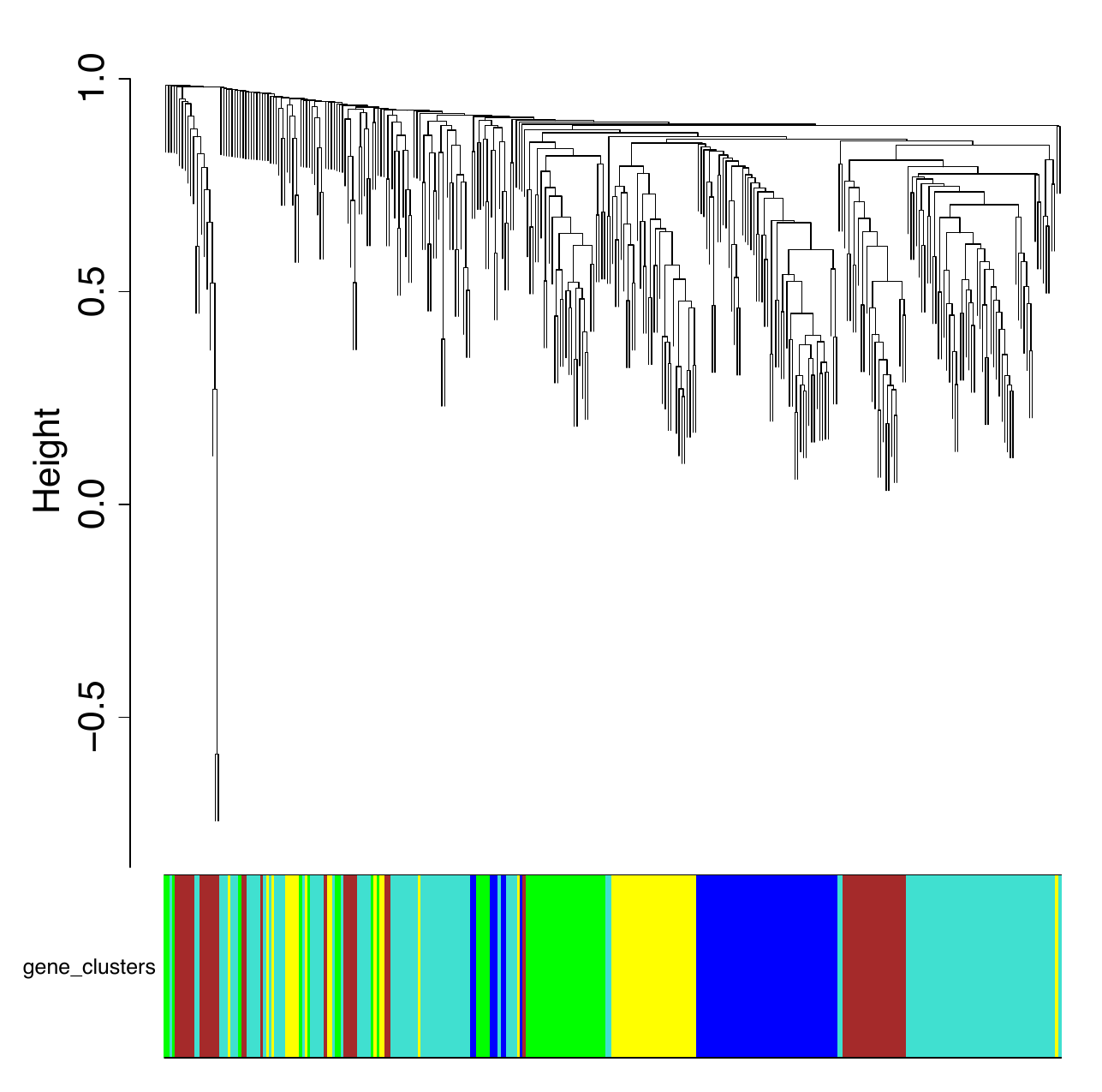}
    \caption{BRCA}\label{fig:dendogram-brca}
  \end{subfigure}%
  \hfill
  \begin{subfigure}{0.32\textwidth}
    \includegraphics[width=\linewidth]{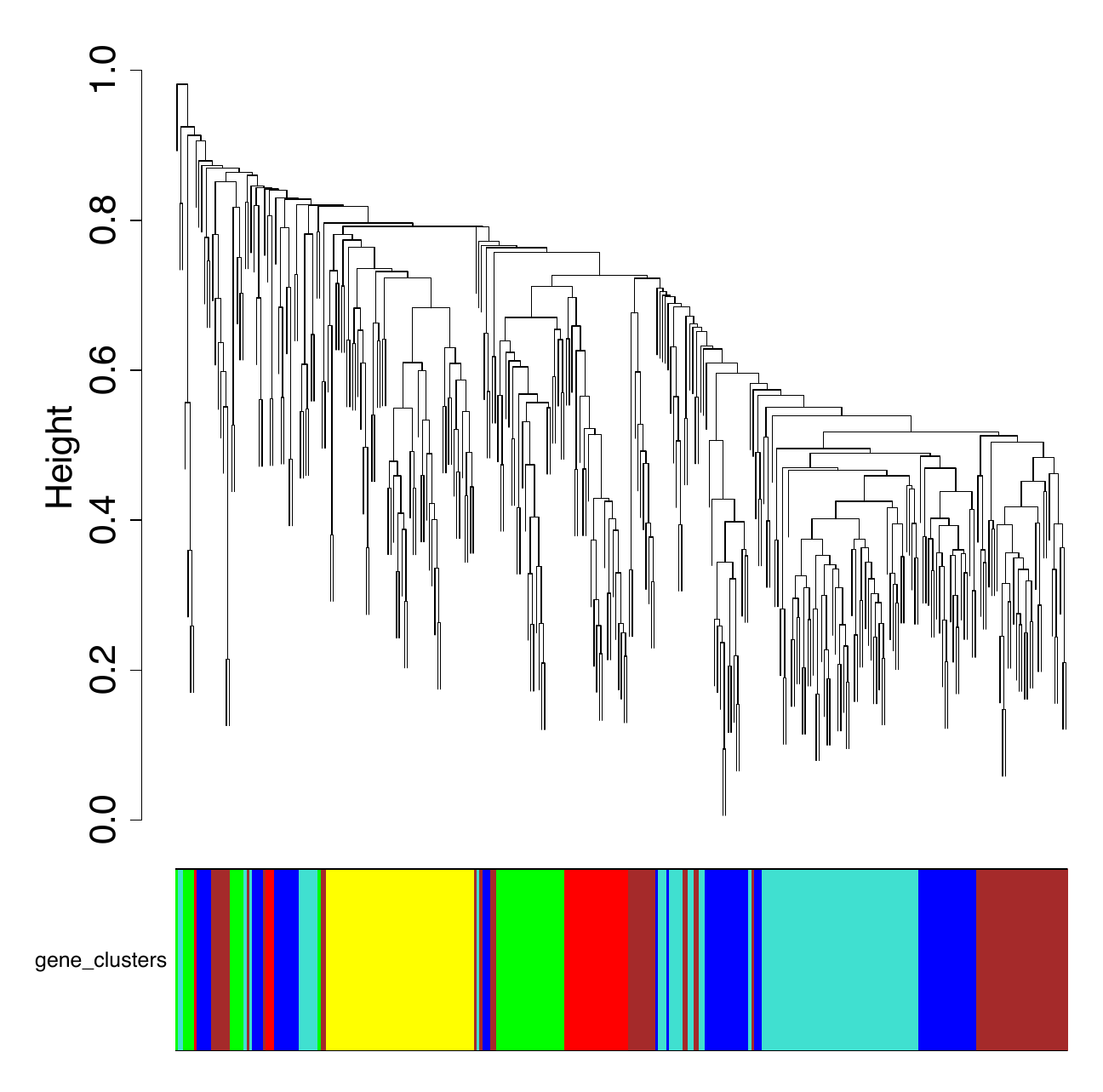}
    \caption{COAD/READ}\label{fig:dendogram-coad}
    \end{subfigure}
      \hfill
  \begin{subfigure}{0.32\textwidth}
    \includegraphics[width=\linewidth]{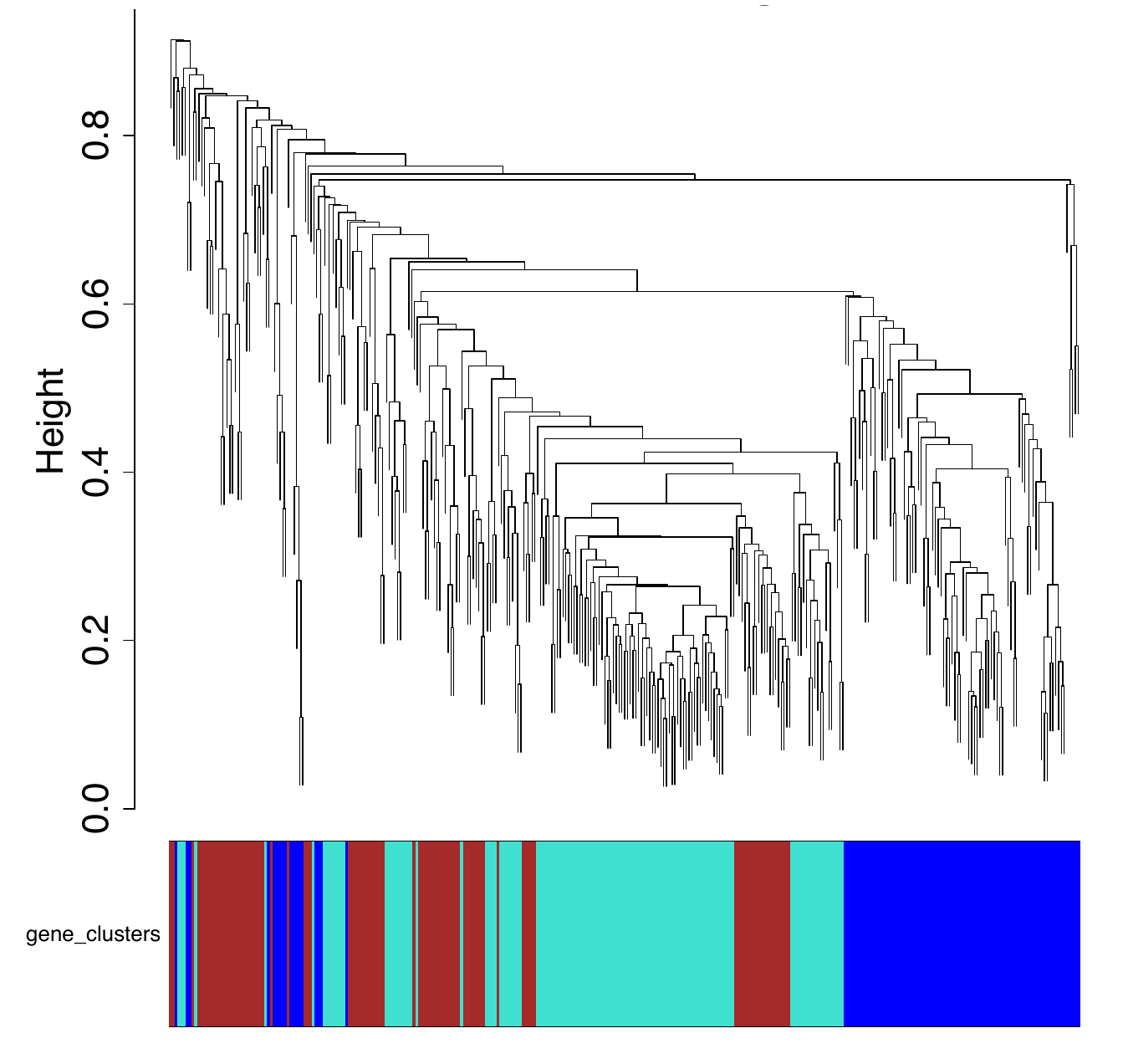}
    \caption{LUAD}\label{fig:dendogram-luad}
  \end{subfigure}%
  \vspace{0.3cm}
  \caption{Gene dendrograms for BRCA, COAD/READ and LUAD gene expression datasets. The colors represent gene modules and the height of the branches indicates the degree of similarity between genes.}\label{fig:dendograms}
\end{figure}

Applying WGCNA to the BRCA, COAD/READ, and LUAD datasets unveiled distinct gene co-expression modules, each marked by unique genes with correlated expression patterns. These are represented by the different colours on the gene dendograms shown in Figure \ref{fig:dendograms}. For BRCA, four unique gene modules were identified, while five were found for COAD/READ, and three for LUAD. Using eigengene-based connectivity (\textit{kME}), hub genes were identified and used as gene signatures for \textit{in-silico} validation. \\

The analysis yielded a total of 60 hub genes: 20 for BRCA, 24 for COAD/READ, and 16 for LUAD. Notable hub genes identified included \textit{PIK3CD, PIK3CG, TRAF1, PIK3R5} for BRCA patients, a prominent gene set recognized in the WGCNA framework, playing crucial roles associated with PI3K signaling critical for cancer cell invasion and metastasis \cite{fruman2014pi3k}. Similarly, \textit{LAMA4, LAMC1, PDGFRB, LAMB2} for COAD/READ patients showed overlaps with gene signatures identified using WGTDA and are associated with liver metastasis in colorectal cancer \cite{steller2013pdgfrb}. Lastly, the gene signature \textit{PDGFRB, JAK1, STAT5B, STAT5A} in LUAD patients, showed associations to signal transducer and activator of transcription (STAT) family of transcription factors, which has recently been implicated as a potential treatment target in lung cancer \cite{faida2023lung}.


\subsubsection{WGTDA}

\begin{figure}[htp!]
\centering

  \begin{subfigure}{0.32\textwidth}
    \includegraphics[width=\linewidth]{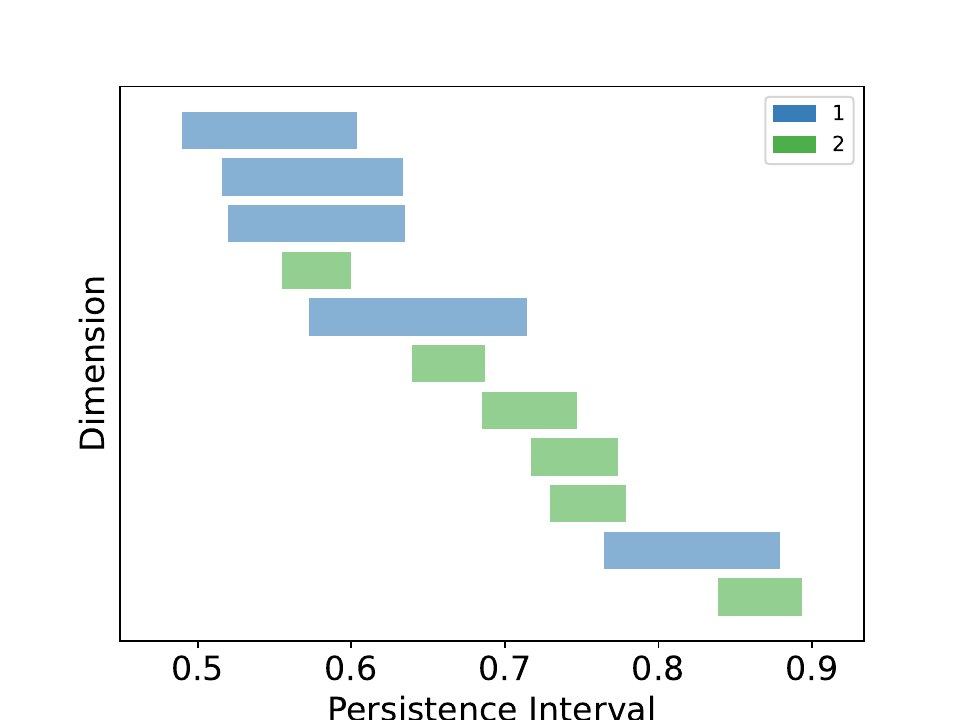}
    \caption{BRCA}\label{fig:barcode-brca}
  \end{subfigure}%
  \hfill
  \begin{subfigure}{0.32\textwidth}
    \includegraphics[width=\linewidth]{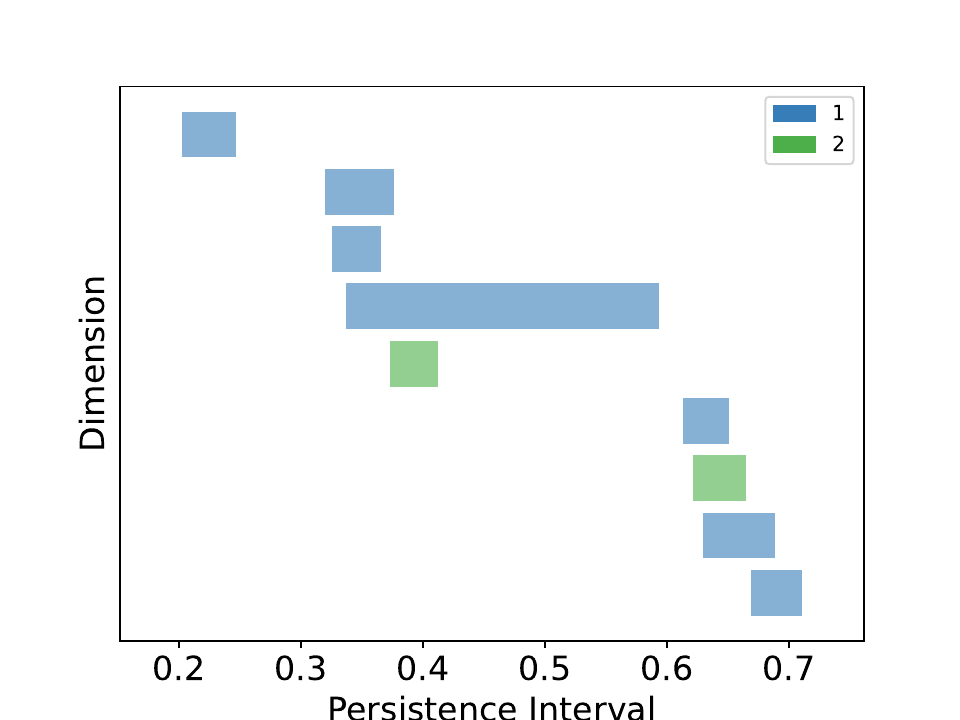}
    \caption{COAD/READ}\label{fig:barcode-coad}
    \end{subfigure}
      \hfill
  \begin{subfigure}{0.32\textwidth}
    \includegraphics[width=\linewidth]{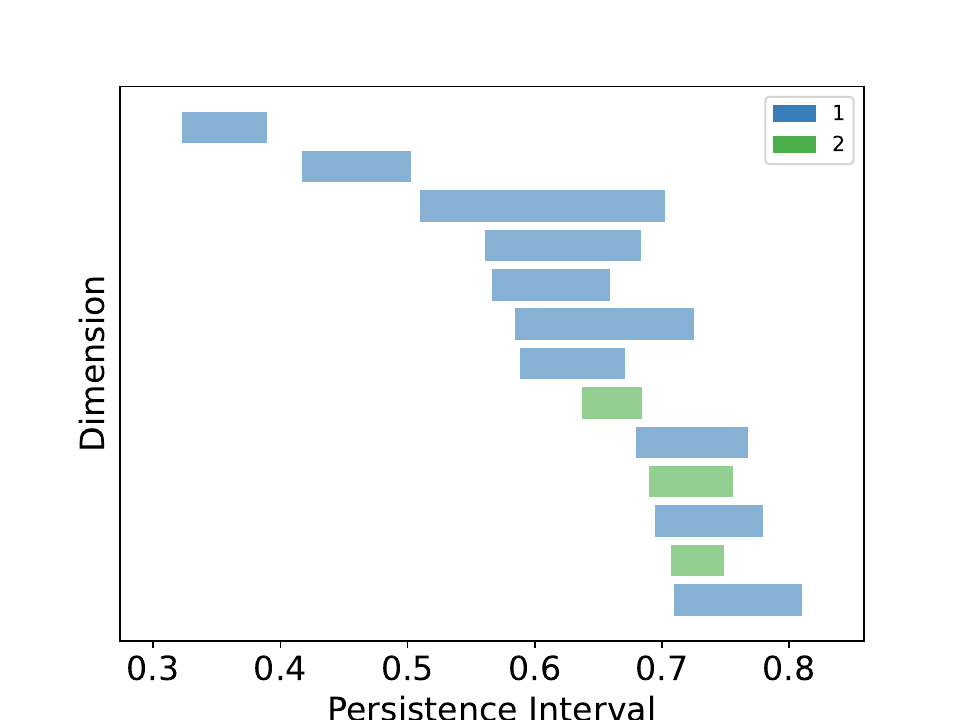}
    \caption{LUAD}\label{fig:barcode-luad}
  \end{subfigure}%
  \vspace{0.4cm}
  \caption{Persistence barcodes for BRCA, COAD/READ, and LUAD gene expression datasets. Each barcode represents the presence and persistence of topological features with \textit{Betti}-1 represented as blue and \textit{Betti}-2 represented as green.}
  \label{fig:barcodes}
\end{figure}

Topological features were revealed in different dimensions by applying WGTDA to BRCA, COAD/READ, and LUAD datasets. A visual summary of the topological features is provided by the persistence barcodes in Figure \ref{fig:barcodes} where blue represents the \textit{Betti}-1 features and green represents the \textit{Betti}-2 features. Notably, the barcodes capture the top 3\% of persistent topological features, which have been selected for \textit{in-silico} validation through survival analysis. For \textit{Betti}-1, five topological features were identified, while seven were found for COAD/READ and ten for LUAD. On the other hand, for \textit{Betti}-2 there were six, two and three topological features found for BRCA, COAD/READ and LUAD, respectively. In contrast to WGCNA, the topological features discovered by WGTDA have duplicate genes for different topological features. The number of gene modules and topological features identified using WGCNA and WGTDA methodologies, with the corresponding number of unique genes for cancer type are shown in Table \ref{tab:sum_of_features}.

\begin{table}[htbp!]
    \centering
    \begin{tabular}{m{2.5cm} m{3.5cm} m{2.5cm} m{2.5cm}}
        \toprule
        \textbf{Cancer Type} & \textbf{WGCNA Modules (\textit{N} genes)} & \multicolumn{2}{c}{\textbf{WGTDA Features (\textit{N} genes)}} \\
        \cmidrule(lr){3-4}
        & & \textbf{\textit{Betti}-1} & \textbf{\textit{Betti}-2} \\
        \midrule
        BRCA & 6 (20) & 5 (13) & 6 (14) \\
        COAD/READ & 8 (24) & 7 (16) & 2 (4) \\
        LUAD & 3 (16) & 10 (20) & 3 (6) \\
        \midrule
        Total & 17 (60)  & 22 (49)  & 11 (24)  \\
        \bottomrule
    \end{tabular}
    
     \vspace{0.3cm}
     \caption{The number of gene modules identified using WGCNA, and number of topological features identified using WGTDA -- including corresponding number of unique genes.}
    \label{tab:sum_of_features}
\end{table}

\subsection{Survival Analysis of Hub Genes and Topological Features}



In this analysis, WGTDA emerged with a higher proportion of significant gene signatures compared to WGCNA for all three cancer datasets. More specifically, the proportion of significant gene signatures as compared to the total amount of gene signatures found were 17.64\% (3 out of 17) for WGCNA, 22.72\% (5 out of 22) for \textit{Betti}-1 (WGTDA framework), and 18.18\% (2 out of 11) for \textit{Betti}-2 (WGTDA framework). Thereby emphasising the potential of WGTDA as a biomarker discovery framework. \\

For the WGCNA framework, where three out of the 17 gene signatures were found to be significant, two of the gene sets were significantly associated with the LUAD and one with BRCA (Table \ref{tab:gene_signatures}). Using the WGTDA framework, \textit{Betti}-1 revealed three gene sets for BRCA, and two for COAD. Lastly, \textit{Betti}-2 revealed one gene signature for BRCA and one for LUAD. Notably, neither the WGCNA, nor the WGTDA-\textit{Betti}-2 frameworks revealed any significant gene signatures for the COAD/READ dataset. In addition, WGTDA-\textit{Betti}-1 had no significant gene signatures for the LUAD dataset. For the complete list of gene signatures identified using the WGCNA and WGTDA frameworks, we direct the reader to the Supplementary Materials. \\



\begin{figure}[!htb]
\centering
  \begin{subfigure}{0.32\textwidth}
    \includegraphics[width=0.95\linewidth]{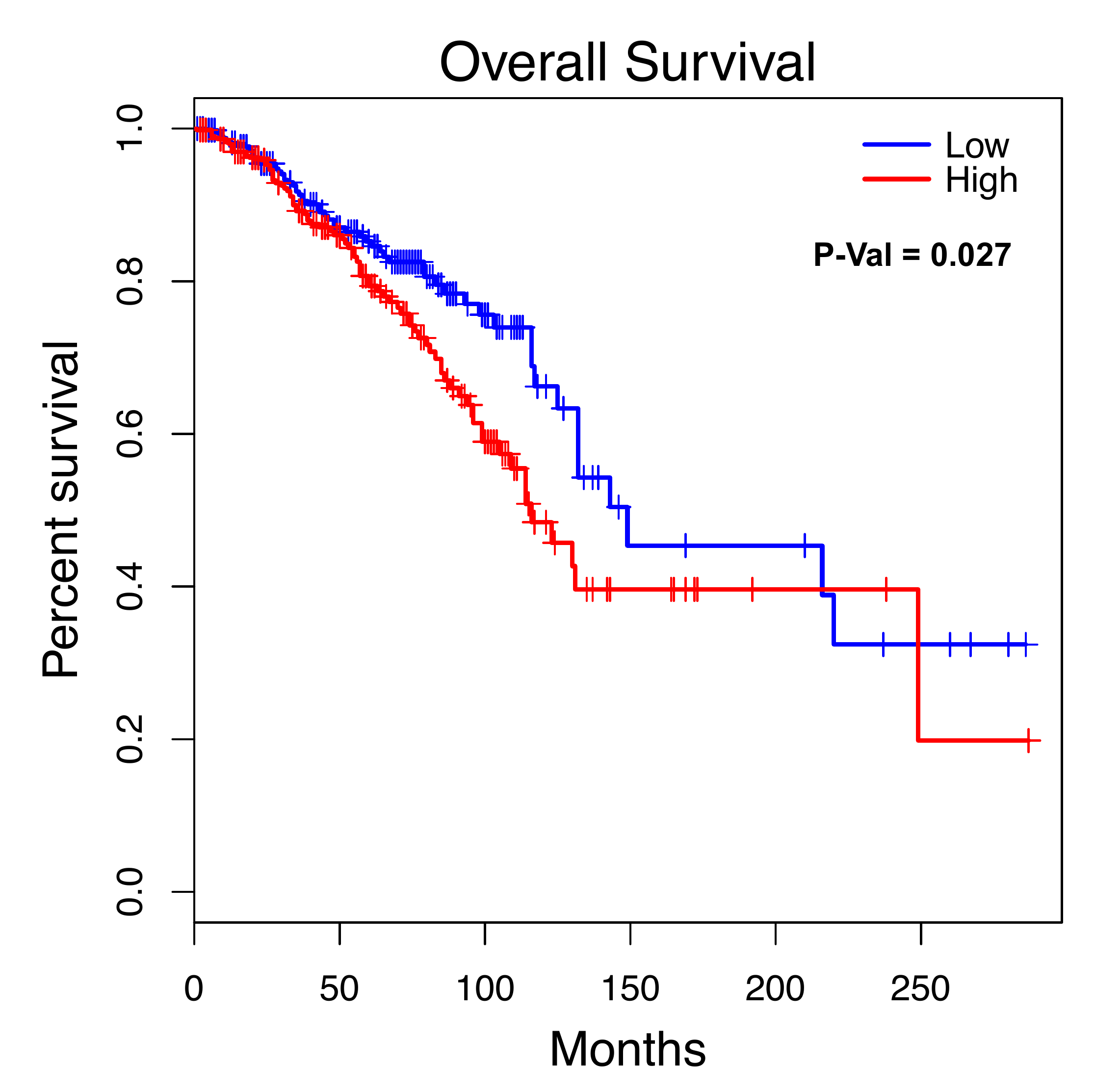}
    \caption{\textit{WGCNA} - BRCA}\label{fig:sa-wgcna-brca}
  \end{subfigure}%
  \begin{subfigure}{0.32\textwidth}
    \includegraphics[width=\linewidth]{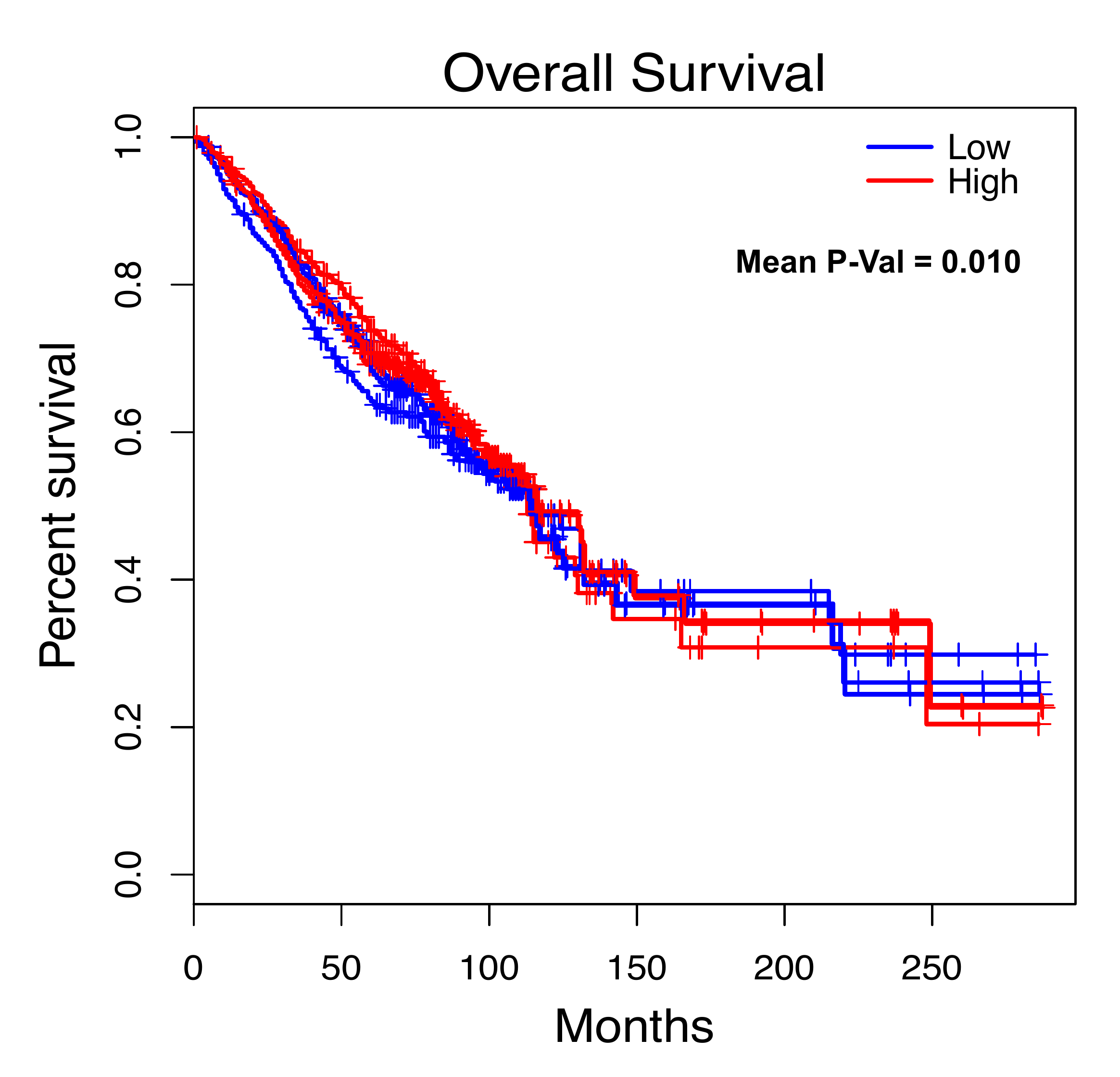}
    \caption{\textit{Betti}-1 - BRCA}\label{fig:sa-b1-brca}
    \end{subfigure}
  \begin{subfigure}{0.32\textwidth}
    \includegraphics[width=\linewidth]{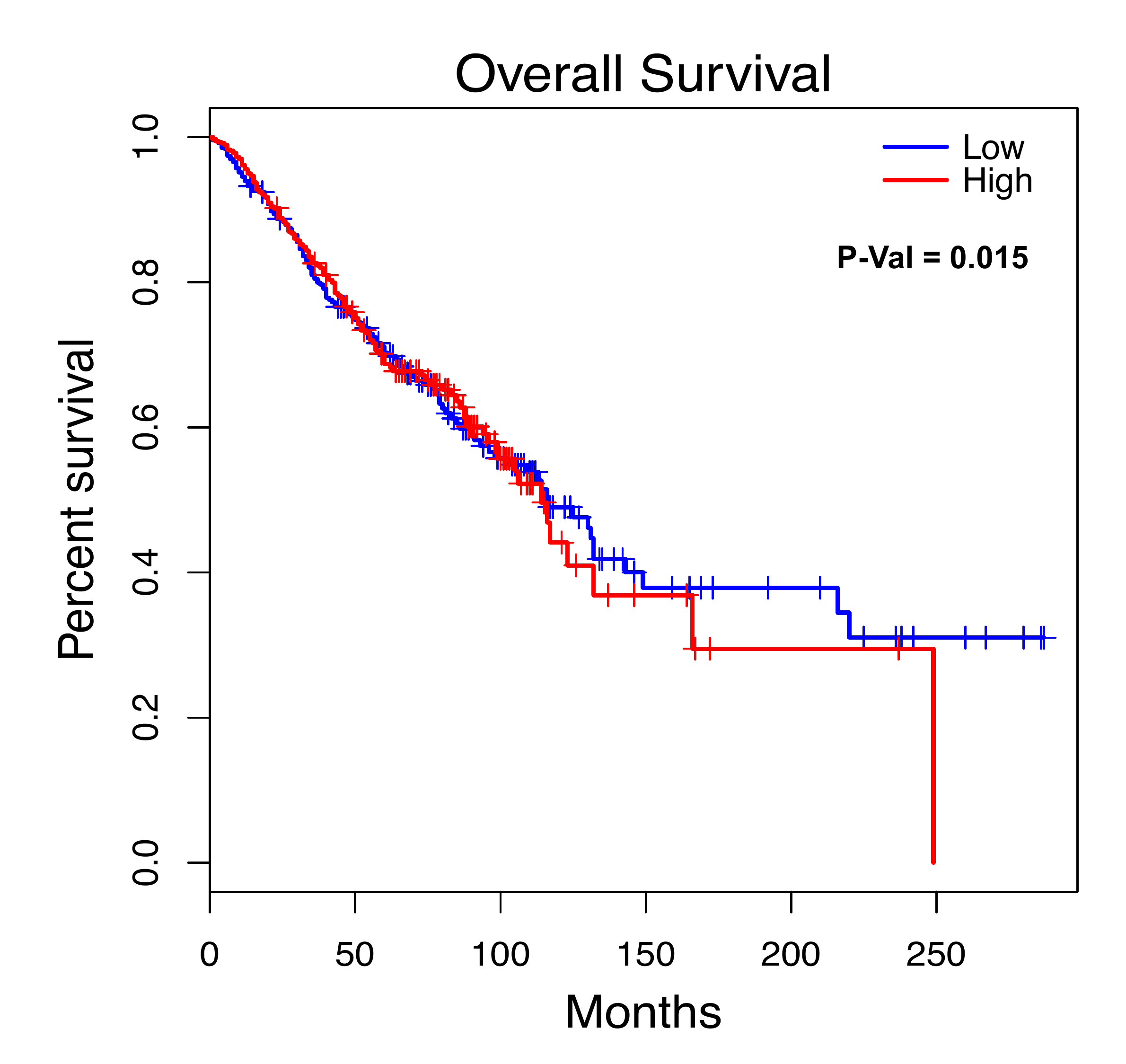}
    \caption{\textit{Betti}-2 - BRCA}\label{fig:sa-b2-brca}
  \end{subfigure}%
  \vspace{0.3cm}
    \centering
  \begin{subfigure}{0.32\textwidth}
    \includegraphics[width=1.03\linewidth]{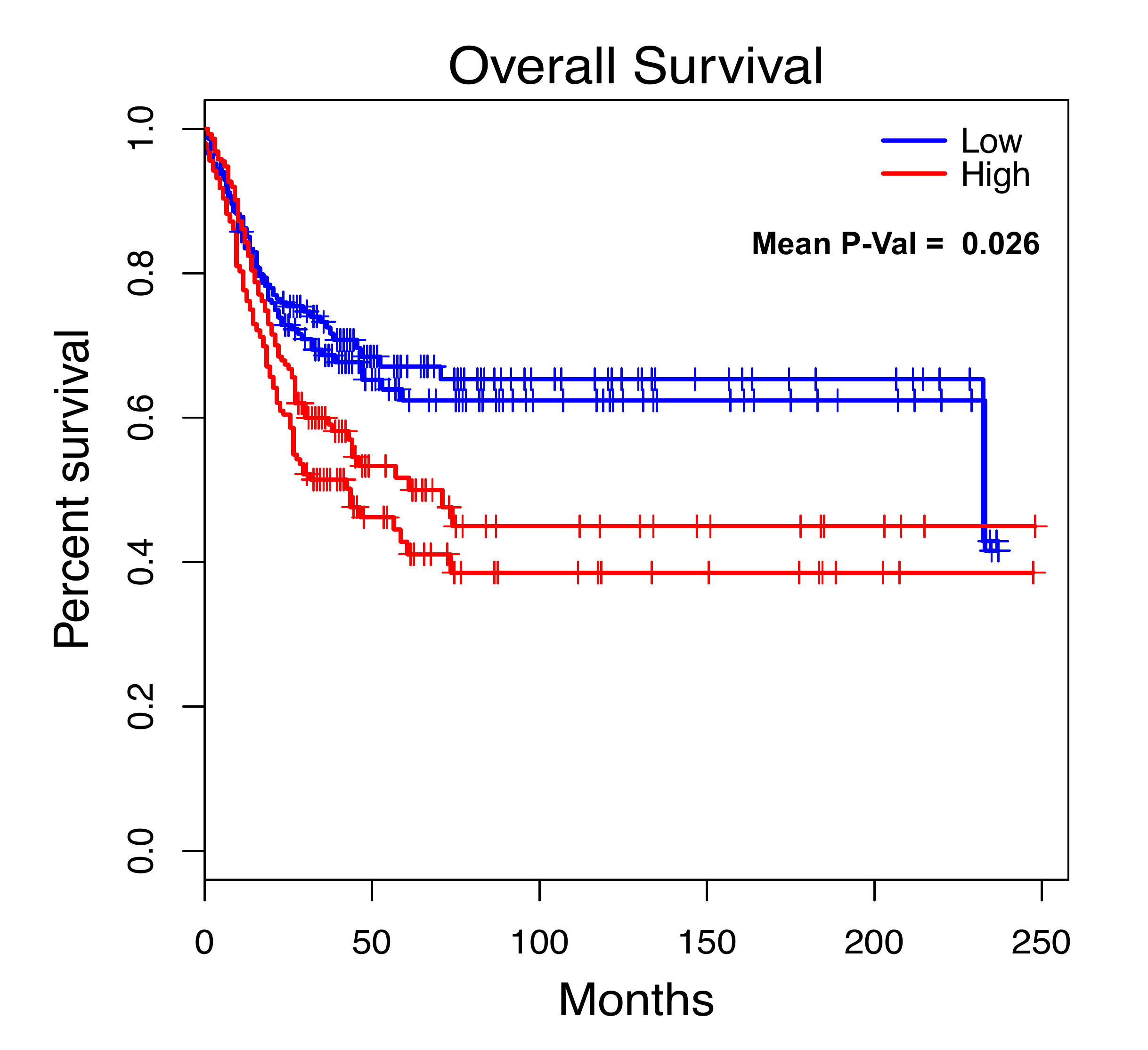}
    \caption{WGCNA - LUAD}\label{fig:sa-wgcna-luad}
    \end{subfigure}
  \begin{subfigure}{0.32\textwidth}
    \includegraphics[width=\linewidth]{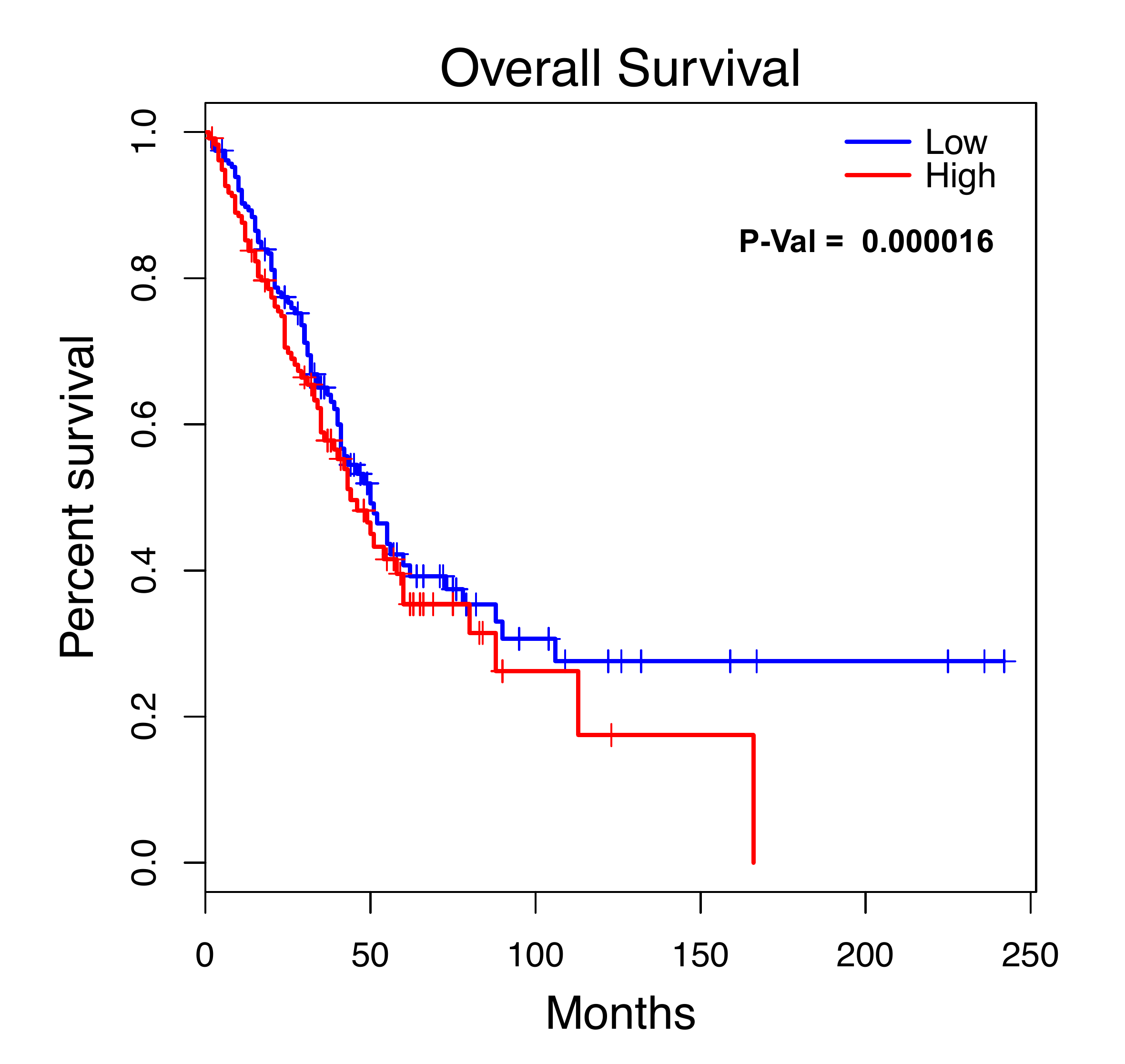}
    \caption{\textit{Betti}-2 - LUAD}\label{fig:sa-b2-luad}
    \end{subfigure}
    \begin{subfigure}{0.33\textwidth}
    \includegraphics[width=\linewidth]{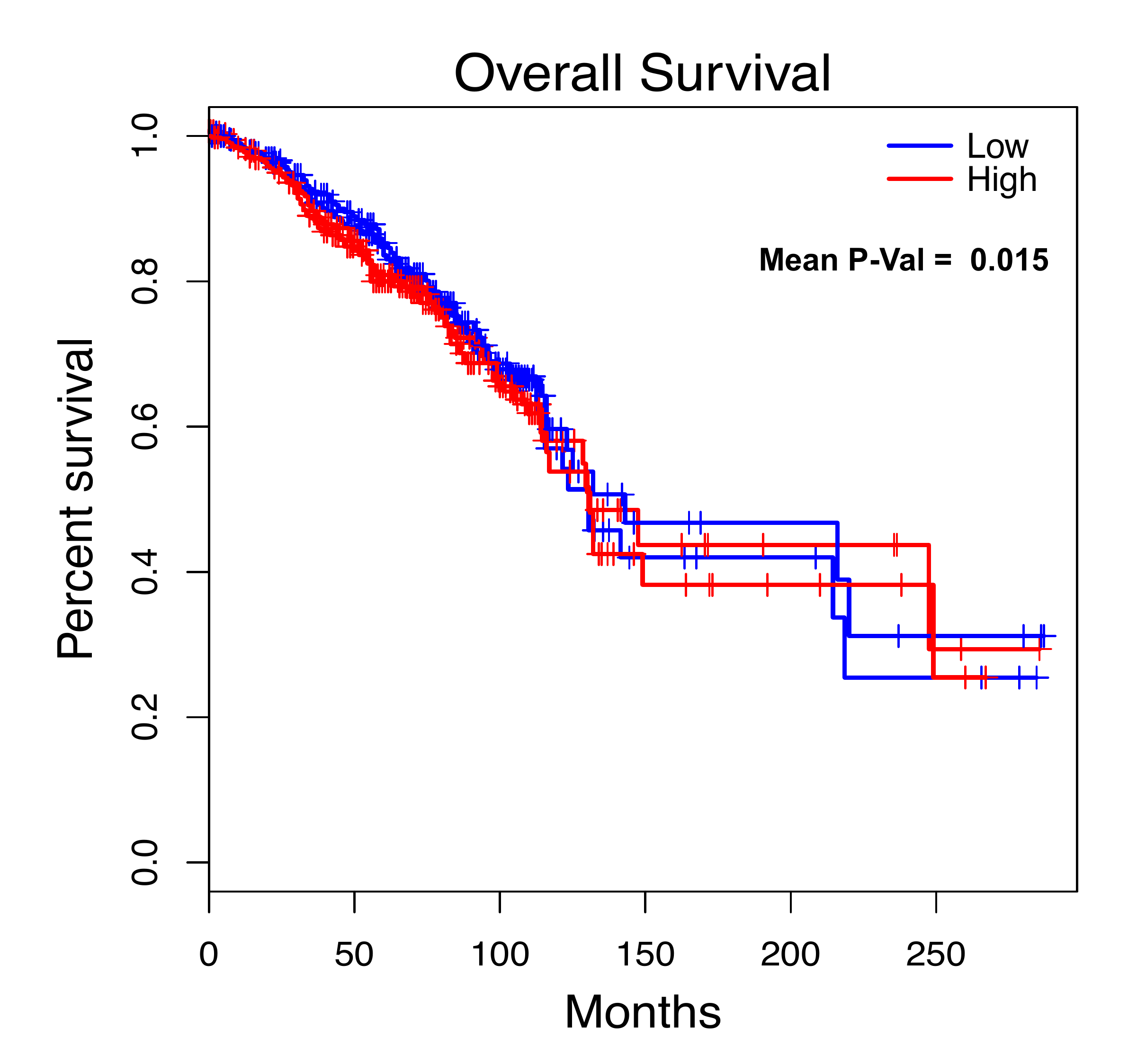}
    \caption{\textit{Betti}-1 - COAD/READ}\label{fig:survival-coad}
    \end{subfigure}
    \vspace{0.3cm}
    \caption{Survival analysis was performed on significant gene signatures in BRCA, COAD/READ, and LUAD datasets. The p-values are reported (mean p-values are reported in cases where multiple significant gene signatures were identified in each method), and the red line indicates high median expression, while the blue line indicates low median expression. }\label{fig:survival}
    \end{figure}

A key finding in the survival analysis, as illustrated in Figures \ref{fig:sa-b1-brca}, \ref{fig:sa-b2-brca}, \ref{fig:sa-b2-luad}, \ref{fig:survival-coad}, showcased that the survival probabilities for the 'high' and 'low' expression groups for \textit{Betti}-1 and \textit{Betti}-2 features, are remarkably close yet maintain statistical significance. This observation is particularly noteworthy as it suggests that the gene signatures identified by WGTDA are significantly correlated with survival outcomes, irrespective of whether the gene expression is below or above the median. Interestingly, the phenomenon is demonstrated for BRCA, COAD/READ and LUAD as depicted in Figures \ref{fig:sa-b1-brca}, \ref{fig:sa-b2-brca}, \ref{fig:survival-coad}, \ref{fig:sa-b2-luad}. The plots show that there is negligible difference between 'high' and 'low' median expression groups, yet these differences are statistically meaningful. The observation from this analysis therefore presented an intriguing hypothesis: WGTDA is identifying sets of gene signatures that are significant to survival probability, regardless of if they are 'low' or 'high' expression. \\


\subsection{Random Survival Forest Analysis}

In this study, three random survival forests were conducted for each cancer dataset to analyse the variable importance of the gene signatures extracted by WGTDA and WGCNA. The random survival forests encompassed all the gene signatures unveiled by its respective dataset. Thus, there were 16 features for BRCA, 15 features for COAD/READ, and 16 features for LUAD, with the target being survival time. The inclusion of all gene signatures for each cancer type enabled a thorough investigation of their relative importance scores. By doing so, a direct comparison and contrast could be performed to assess the impact of WGTDA and WGCNA gene signatures within the same predictive model. Gene signatures' variable importance scores are shown in Table \ref{tab:gene_signatures} \\

In the RSF analysis, notable findings in the variable importance scores amongst the gene signatures were identified by WGCNA and WGTDA across the various cancer types (Refer to Supplementary Section for all gene signatures variable important scores). For BRCA and LUAD, the WGTDA method stood out yielding high variable importance scores for the gene signatures as compared to WGCNA. This observation is particularly evident in the case of BRCA with the gene signature \textit{B1\_Signature\_1}, and in LUAD with the gene signatures \textit{B1\_Signature\_7} and \textit{B2\_Signature\_3}. On the other hand, we did not observe any significant gene signatures for the COAD/READ cohort using both methods. Furthermore, \textit{Betti}-1 performed better than WGCNA for BRCA and LUAD cohorts, with \textit{Betti}-1\ performing the best on the BRCA cohort. While \textit{Betti}-2 performed better than WGCNA on the LUAD cohort. \\

 To further investigate this phenomenon, we follow with a discussion on the biological relevance of these significant \textit{Betti} numbers. This exploration will be conducted through functional enrichment analysis. This is important as as it moves beyond simply identifying statistically significant patterns towards interpreting what these patterns mean in a biological context. \\



\begin{table}
    \centering

    \begin{tabular}{@{}llcc@{}}
        \toprule
        Signature & Genes & Survival $P$-Value & Importance Ratio \% \\
        \midrule
        WGCNA\_Signature\_1\_BRCA & \textit{XIAP, MAPK1, APPL1, CHUK} & 0.027 & 3.9\\
        B1\_Signature\_1\_BRCA & \textit{SPI1, PRKCB, RAC2} & 0.025 & 4.8 \\
        B1\_Signature\_2\_BRCA & \textit{MSH3, APC, PIAS1} & 0.0058 & 4.8 \\
        B1\_Signature\_5\_BRCA & \textit{PIAS1, APC, SOS2} & 0.0093 & 4.8 \\
        B2\_Signature\_5\_BRCA & \textit{MSH3, APC, APC} & 0.0015 & 2.1  \\
        
        B1\_Signature\_1\_COAD & \textit{MAPK1, CRKL} & 0.0069 & 1.6 \\
        B1\_Signature\_7\_COAD & \textit{LAMC1, LAMA4} & 0.017  & 0.0 \\

        WGCNA\_Signature\_1\_LUAD & \textit{PDGFRB, JAK1, STAT5B, STAT5A} & 0.002 & 2.2 \\
        WGCNA\_Signature\_2\_LUAD & \textit{RELA, PIAS4, CTBP1, RXRB} & 0.05 & 2.2 \\
        B2\_Signature\_3\_LUAD & \textit{BIRC5, RAD51} & 0.000016 & 5.1 \\
        \bottomrule
    \end{tabular}
    \vspace{0.3cm}
     \caption{Gene signatures identified using WGCNA and WGTDA with survival p-value and importance score. }
     \label{tab:gene_signatures}
\end{table}
\subsection{Functional Enrichment}

Functional enrichment analyses were conducted on gene signatures identified through WGCNA and WGTDA, particularly those significantly associated ($p$-value $<$ 5\%) with poor patient prognosis (Refer to Supplementary Section). Notable insights from WGCNA hub genes revealed a connection between RHO GTPases activation of NADPH oxidases, linking cellular signaling to reactive oxygen species (ROS) production \cite{lin2023rho}. WGTDA underscored ROS production in BRCA patients. In COAD/READ patients, WGCNA hub genes implicated abnormal activation of the epidermal growth factor receptor (EGFR), involving constitutive signaling by EGFRvIII and ligand-responsive EGFR cancer variants pathways. Conversely, WGTDA pointed towards the activation of the FGFR3 receptor, particularly through aberrant ligand binding, commonly associated with digestive tract cancers and multiple myeloma \cite{xiao2021targetable}. Furthermore, WGCNA hub genes in LUAD patients involved IL-21, IL-5, and IL-15 pathways, influencing immune cell activity within the tumor microenvironment \cite{zhang2021biological}. WGTDA highlighted pathways such as TP53 signaling in regulating transcription of cell death genes and the impairment of BRCA2 binding to PALB2, disrupting DNA repair interactions and increasing the risk of genomic instability \cite{sadeghi2020molecular}. Collectively, these findings provide a comprehensive understanding of how biological pathways identified by both WGCNA and WGTDA collectively modulate cellular signaling, apoptotic regulation, and redox dynamics in cancer.

\section{Discussion} \label{sec:discussion}
This study presents WGTDA, a novel framework to identify biomarkers rooted in topology and persistent homology principles. The analysis revealed that WGTDA not only identified a greater proportion of significant gene signatures than WGCNA but also demonstrated a higher variable importance suggesting that \textit{Betti} features may be predictive of survival outcomes. Moreover, WGTDA also revealed a distinct pattern in the data suggesting a level of certainty in predicting premature survival outcomes. This pattern, observed in Kaplan-Meier model in Figure \ref{fig:survival}, showcases the potential of WGTDA as a highly promising and impactful tool in the field of genomic research. \\

A key observation was the close proximity between 'high' and 'low' expression levels for significant gene signatures in predicting survival outcomes. The implication of this pattern may be profound, suggesting that the gene signatures identified by WGTDA are robust indicators of survival probability, regardless of whether the gene expression level is above or below the median. This consistency in survival outcomes, irrespective of the gene expression's high or low status, underlines the precision of WGTDA in capturing crucial survival predictors. It also hints at a deeper, more intricate interplay of genetic factors in influencing cancer prognosis, one that might not be solely dependent on the magnitude of gene expression but also on their topological and network properties within the genomic landscape. \\

Furthermore, when comparing the results from the RSF model and the Kaplan-Meier model, it was found that gene signatures exhibiting the highest variable importance in the RSF models also showed minimal differences in the median high and low expression levels, yet these differences were statistically significant in terms of survival outcomes (Table \ref{tab:gene_signatures}). This is demonstrated in gene signature \textit{B2\_Signature\_3} in the Supplementary Section, with the corresponding KM plot being Figure \ref{fig:sa-b2-luad}. The consistency of this pattern across both KM and RSF models attests to its significance. In our examination of \textit{B2\_Signature\_3}, we observed that the gene set includes \textit{BIRC5} and \textit{RAD51}. Particularly, the overexpression of \textit{BIRC5} has recently been shown to play a significant role in modulating lung cancer stem cells and activating epithelial-to-mesenchymal transition (EMT) \cite{kahm2023birc5}. Both of these processes are strongly linked to drug resistance, relapse, and metastasis in cancer \cite{phi2018cancer}. Moreover, previous studies have implicated the expression of \textit{RAD51} in enhancing DNA damage repair and promoting survival in lung cancer cells. This suggests that targeting both \textit{BIRC5} and \textit{RAD51} may be an effective therapeutic strategy to overcome drug resistance, especially in \textit{KRAS}-mutant cancers, including lung cancer \cite{hu2019high}. This convergence of results from both KM and RSF analyses emphasises the robustness of these topological-based gene signatures as key indicators in survival prediction, thus affirming their significance as key indicators to patient prognosis. \\

Future experiments to validate and enhance WGTDA's utility include \textit{in vitro} and \textit{in vivo} studies for affirming identified gene signatures in real biological contexts. Moreover, quantum computing systems can expedite calculations of higher-order \textit{Betti} numbers (\textit{Betti}-3, -4, -5), revealing more intricate gene interactions and potential cancer biomarkers. Emphasizing WGTDA as a novel framework highlighting its innovation in biomarker discovery, attracting attention and encouraging broader adoption.

\newpage
\printbibliography

\appendix
\section*{Supplementary Materials}

\subsection*{Supplementary Table 1: Gene Signatures for BRCA}

\begin{longtable}{ll}
\toprule
\textbf{Signature Set} & \textbf{Genes} \\
\midrule
\endhead
\bottomrule
\endfoot
WGCNA{\_}Signature{\_}1 & \textit{XIAP, MAPK1, APPL1, CHUK} \\ 
WGCNA{\_}Signature{\_}2 & \textit{PIK3CD, PIK3CG, TRAF1, PIK3R5} \\ 
WGCNA{\_}Signature{\_}3 & \textit{AXIN1, PIAS4, CTBP1, MAP2K2} \\ 
WGCNA{\_}Signature{\_}4 & \textit{FOXO1, LAMC1, EPAS1, LAMA4} \\ 
WGCNA{\_}Signature{\_}5 & \textit{MSH2, MSH6, CKS1B, RAD51} \\

Betti-1{\_}Signature{\_}1 & \textit{SPI1, PRKCB, RAC2} \\
Betti-1{\_}Signature{\_}2 & \textit{MSH3, APC, PIAS1} \\
Betti-1{\_}Signature{\_}3 & \textit{RAD51, E2F2}  \\
Betti-1{\_}Signature{\_}4 & \textit{E2F1, RAD51} \\ 
Betti-1{\_}Signature{\_}5 & \textit{PIAS1, APC, SOS2} \\ 

Betti-2{\_}Signature{\_}1 & \textit{PRKCB, PIK3CG, PRKCB} \\
Betti-2{\_}Signature{\_}2 & \textit{PIK3CD, PRKCB} \\ 
Betti-2{\_}Signature{\_}3 & \textit{PIK3R5, PIK3CG} \\
Betti-2{\_}Signature{\_}4 & \textit{PIK3CG, PIK3CD, TRAF1} \\
Betti-2{\_}Signature{\_}5 & \textit{MSH3, APC, APC} \\ 
Betti-2{\_}Signature{\_}6 & \textit{TRAF1, PIK3R5, SPI1} \\ 
\end{longtable}

\subsection*{Supplementary Table 2: Gene Signatures for COAD/READ}

\begin{longtable}{ll}
\toprule
\textbf{Signature Set} & \textbf{Genes} \\
\midrule
\endhead
\bottomrule
\endfoot
WGCNA{\_}Signature{\_}1 & \textit{CUL2, RHOA, TPM3, CHUK} \\ 
WGCNA{\_}Signature{\_}2 & \textit{PIAS4, RELA, RXRB, DVL3} \\ 
WGCNA{\_}Signature{\_}3 & \textit{LAMA4, LAMC1, PDGFRB, LAMB2} \\ 
WGCNA{\_}Signature{\_}4 & \textit{RASSF5, SPI1, PIK3R5, PIK3CD} \\ 
WGCNA{\_}Signature{\_}5 & \textit{MAPK1, JAK1, SOS1, RAF1} \\ 
WGCNA{\_}Signature{\_}6 & \textit{FZD4, ARNT, PDGFRA, LAMC1} \\ 

Betti-1{\_}Signature{\_}1 & \textit{MAPK1, CRKL} \\ 
Betti-1{\_}Signature{\_}2 & \textit{PIK3CA, SOS1} \\ 
Betti-1{\_}Signature{\_}3 & \textit{EP300, CREBBP} \\ 
Betti-1{\_}Signature{\_}4 & \textit{CSF1R, SPI1} \\ 
Betti-1{\_}Signature{\_}5 & \textit{CREBBP, SOS1, CBL} \\ 
Betti-1{\_}Signature{\_}6 & \textit{CBL, SOS1, PIAS1} \\ 
Betti-1{\_}Signature{\_}7 & \textit{LAMC1, LAMA4} \\ 

Betti-2{\_}Signature{\_}1 & \textit{SOS1, EP300} \\ 
Betti-2{\_}Signature{\_}2 & \textit{FGF16, VEGFD} \\ 
\end{longtable}

\subsection*{Supplementary Table 3: Gene Signatures for LUAD}

\begin{longtable}{ll}
\toprule
\textbf{Signature Set} & \textbf{Genes} \\
\midrule
\endhead
\bottomrule
\endfoot
WGCNA{\_}Signature{\_}1 & \textit{PDGFRB, JAK1, STAT5B, STAT5A} \\ 
WGCNA{\_}Signature{\_}2 & \textit{RELA, PIAS4, CTBP1, RXRB} \\ 
WGCNA{\_}Signature{\_}3 & \textit{SOS1, APPL1, RAF1, MLH1} \\ 

Betti-1{\_}Signature{\_}1 & \textit{MAPK1, CRKL} \\ 
Betti-1{\_}Signature{\_}2 & \textit{EP300, CREBBP} \\ 
Betti-1{\_}Signature{\_}3 & \textit{APPL1, RAF1} \\ 
Betti-1{\_}Signature{\_}4 & \textit{SMAD2, SMAD4} \\ 
Betti-1{\_}Signature{\_}5 & \textit{RAF1, MLH1} \\ 
Betti-1{\_}Signature{\_}6 & \textit{PIK3CA, Signature-K3B} \\
Betti-1{\_}Signature{\_}7 & \textit{E2F1, E2F2} \\ 
Betti-1{\_}Signature{\_}8 & \textit{MTOR, TPR} \\ 
Betti-1{\_}Signature{\_}9 & \textit{SOS1, Signature-K3B} \\ 
Betti-1{\_}Signature{\_}10 & \textit{PDGFRB, LAMA4} \\ 

Betti-2{\_}Signature{\_}1 & \textit{MSH6, MSH4} \\ 
Betti-2{\_}Signature{\_}2 & \textit{MAP2K2, PIAS4} \\ 
Betti-2{\_}Signature{\_}3 & \textit{BIRC5, RAD51} \\ 
\end{longtable}

\subsection*{Supplementary Figure 1: Importance Scores for all gene signatures and the Associated Functional Enrichment Dotplot For Significant Gene Signatures.}
\setcounter{figure}{0}
\begin{figure}[!htb]
\centering
  \begin{subfigure}{0.47\textwidth}
    \includegraphics[width=\linewidth]{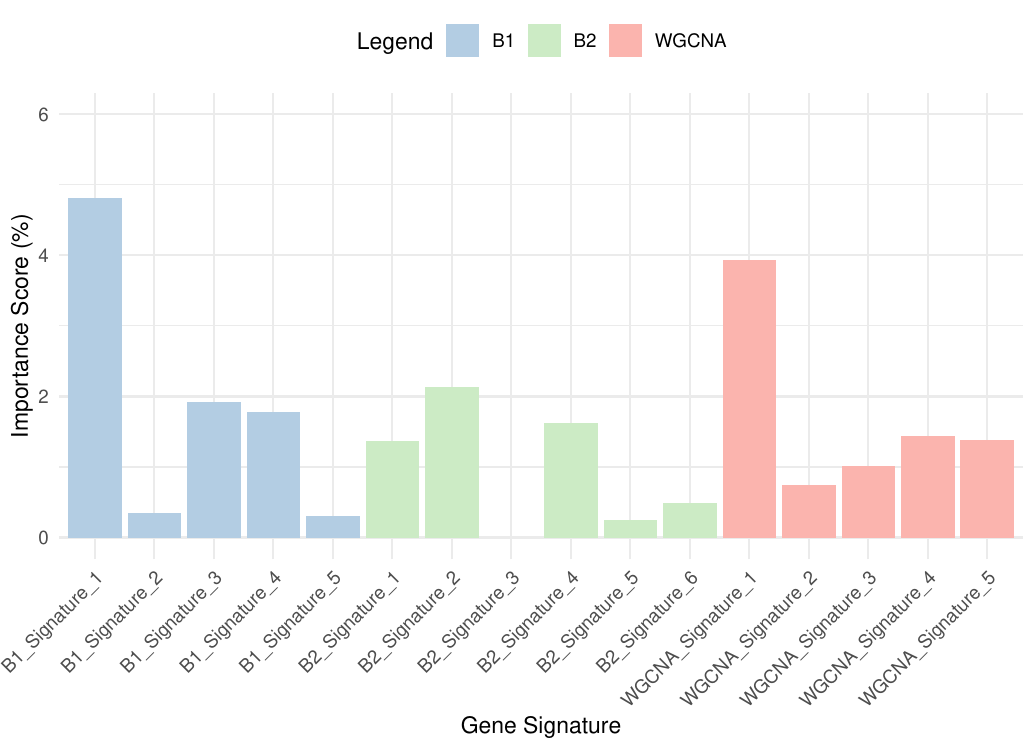}
    \caption{BRCA}\label{fig:vatr_imp-brca}
  \end{subfigure}%
   \hfill
  \begin{subfigure}{0.47\textwidth}
    \includegraphics[width=\linewidth]{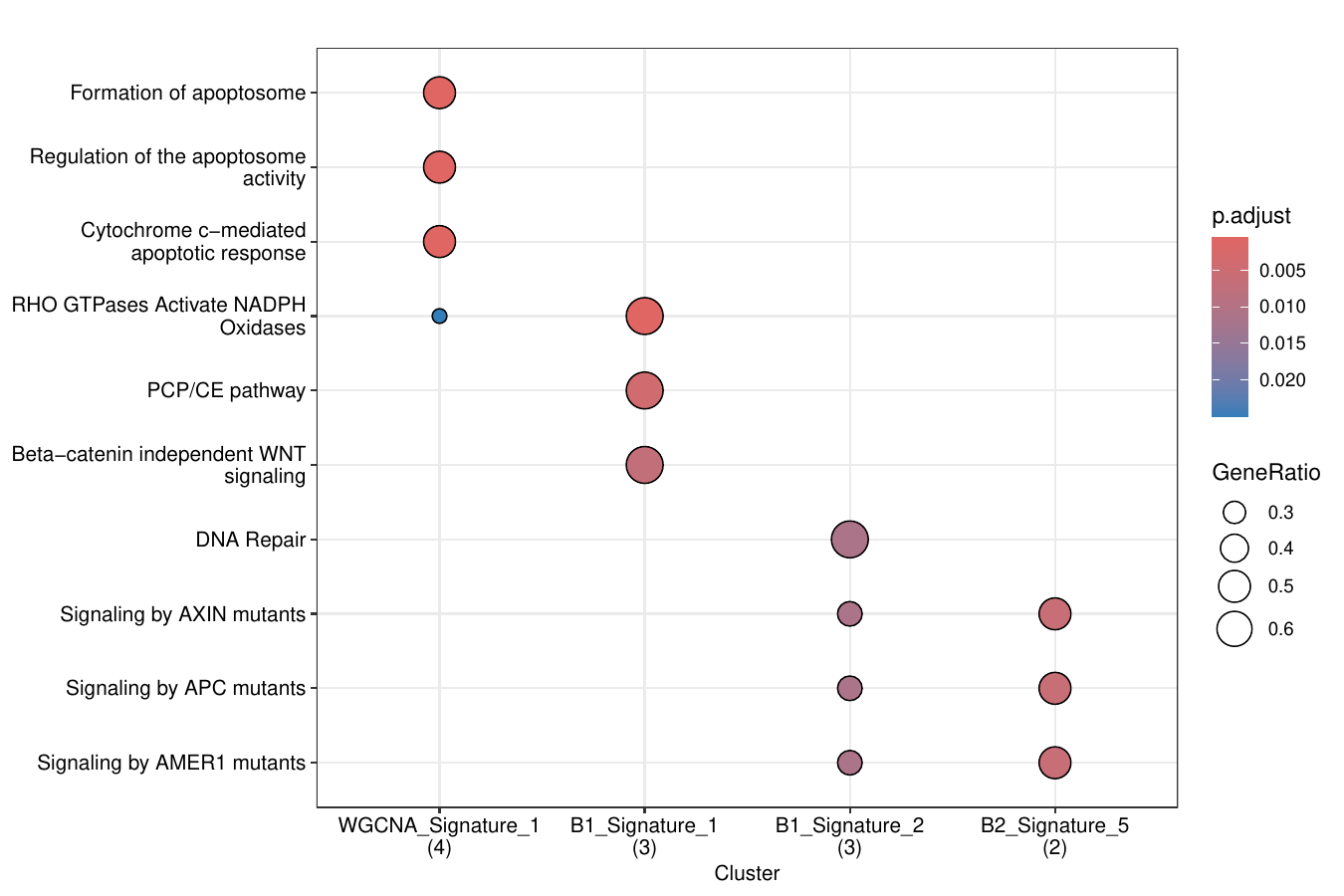}
    \caption{BRCA}\label{fig:reactome-brca}
  \end{subfigure}%
   
  \begin{subfigure}{0.47\textwidth}
    \includegraphics[width=\linewidth]{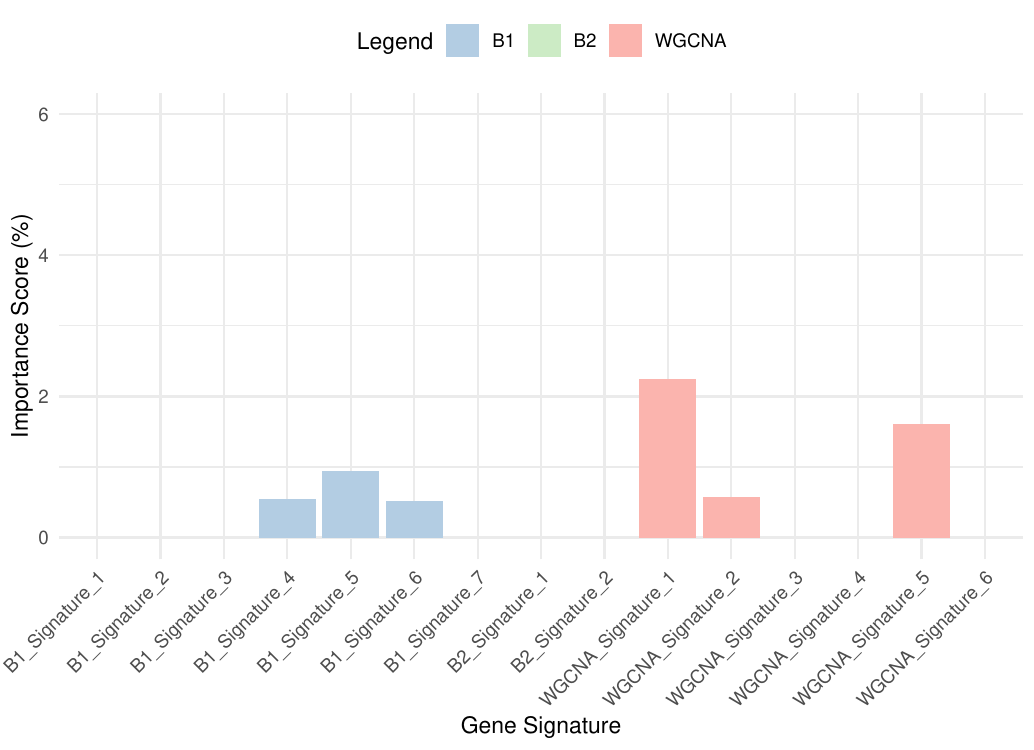}
    \caption{COAD/READ}\label{fig:vatr_imp_coad}
  \end{subfigure}
  \begin{subfigure}{0.47\textwidth}
      \includegraphics[width=\linewidth]{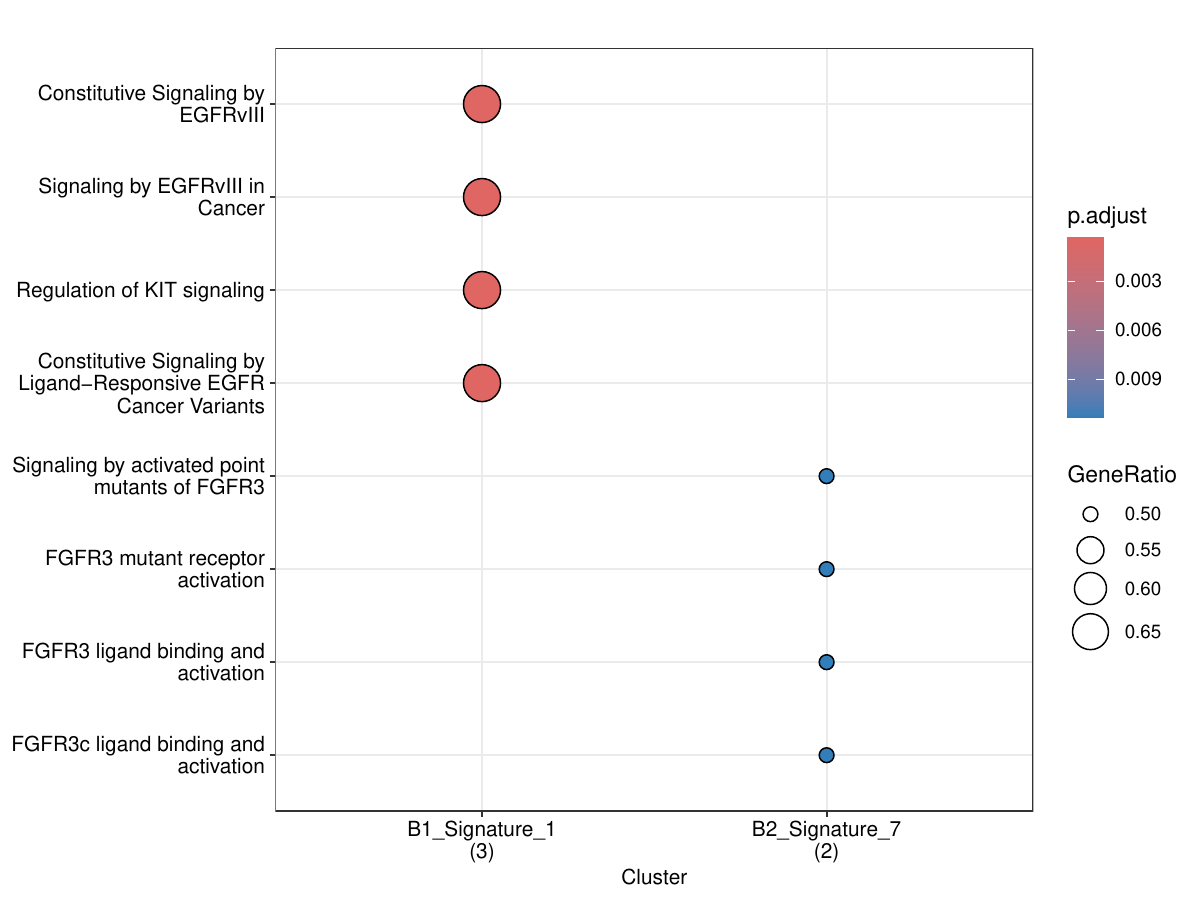}
      \caption{COAD/READ}\label{fig:reactome-coad}
  \end{subfigure}
\end{figure}
\begin{figure}[htbp!]\ContinuedFloat
    \begin{subfigure}{0.47\textwidth}
    \includegraphics[width=\linewidth]{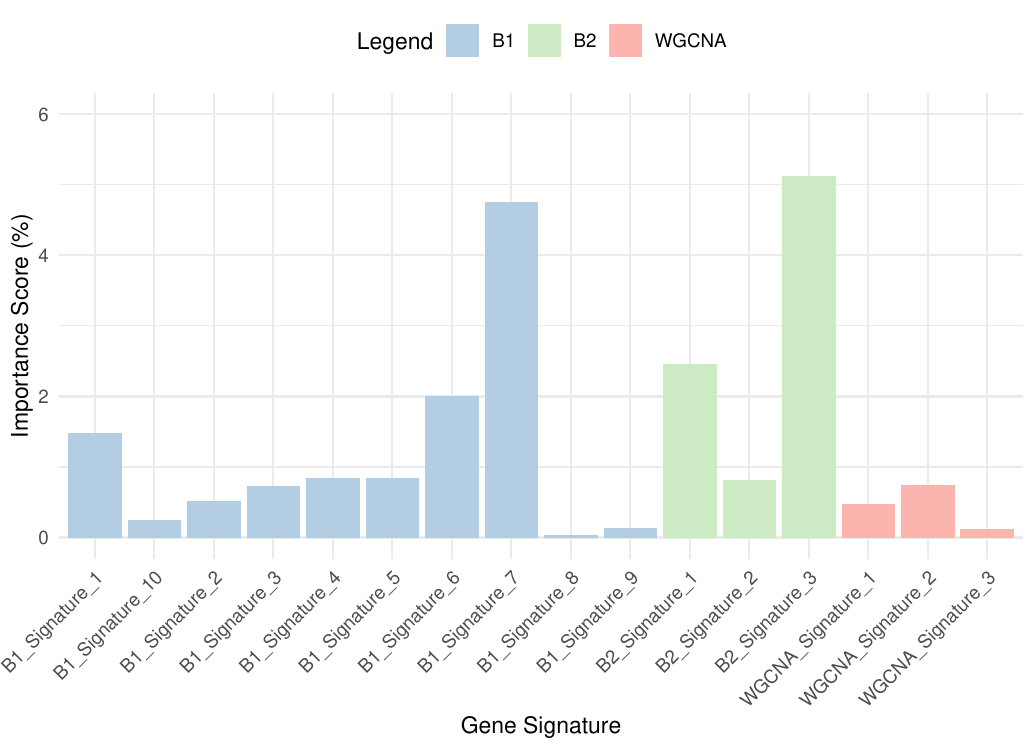}
    \caption{LUAD}\label{fig:vatr_imp_luad}
  \end{subfigure}
  \begin{subfigure}{0.47\textwidth}
      \includegraphics[width=\linewidth]{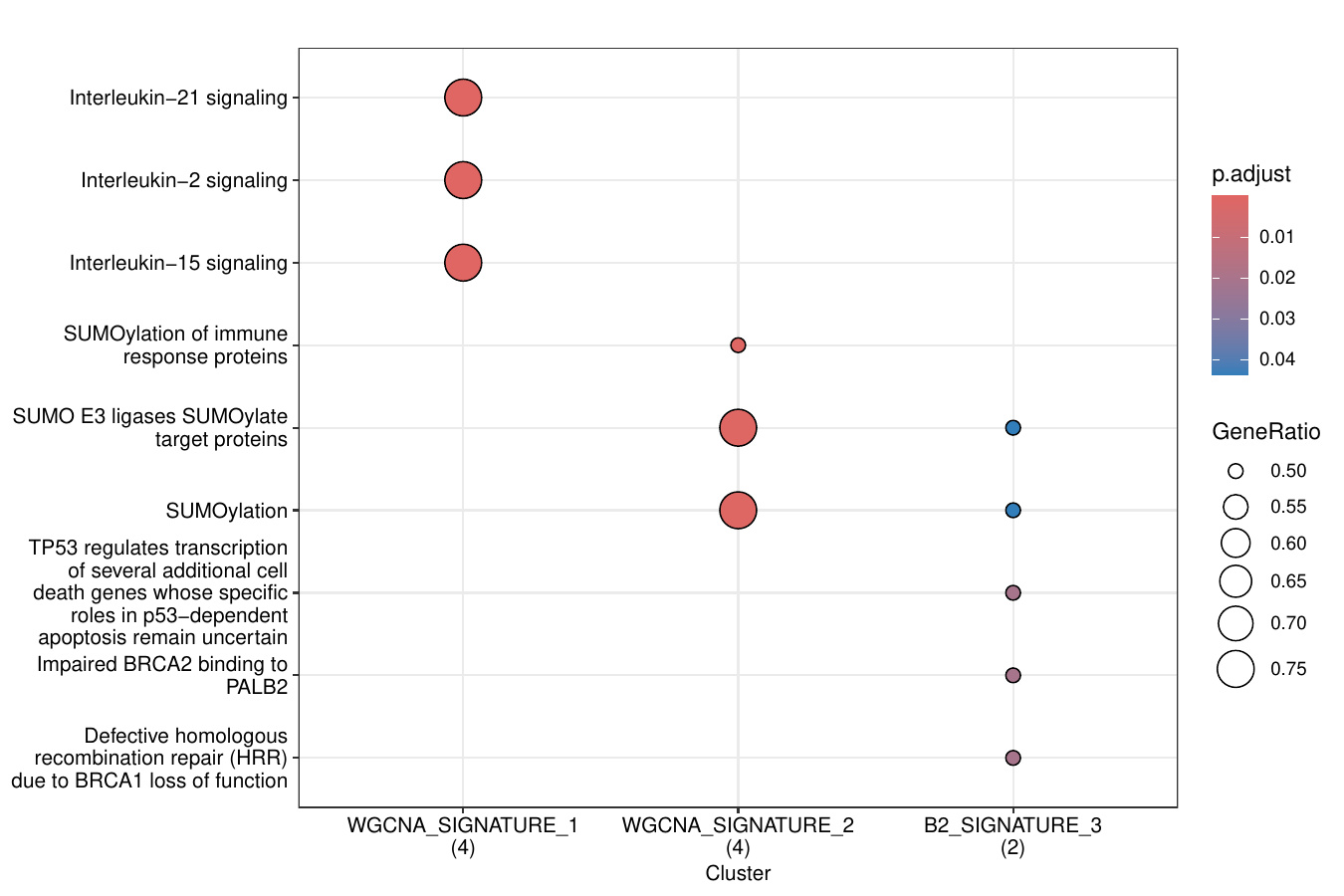}
      \caption{LUAD}\label{fig:reactome-luad}
  \end{subfigure}
  \vspace{0.4cm}
  \caption{The feature importance score bar plot for identified gene signatures using WGCNA and WGTDA and the associated functional enrichment dotplot. The enriched Reactome pathways associated with gene signatures of interest using TCGA BRCA, COAD/READ and LUAD gene expression datasets are shown. Each point's color corresponds to the BH adjusted p-value, while the size reflects the gene ratio. The numbers on the x-axis indicate the count of unique input genes used. }\label{fig:var_enrich}
\end{figure}

\end{document}